\documentclass[12pt]{article}
\pdfoutput=1
\usepackage{graphicx,epstopdf,amssymb,amsfonts,amsmath,amsthm,array,
mathrsfs,amscd}
\newcommand{\pbs}[1]{\let\temp=\\#1\let\\=\temp}
\numberwithin{equation}{section}
\def\be{\begin{equation}}\def\ee{\end{equation}}
\def\cvp{\raise 2pt\hbox{,}} 
 \def\tr{\mathop{\rm tr}\nolimits}

 \def\d{{\rm d}}\def\nn{{\cal
N}} 
 \def\uN{\text{U}(N)}

\def\gs{g_{\text s}}\def\ls{\ell_{\text s}}\def\apr{\alpha'}
\def\AdS{\text{AdS}_{5}}\def\Sfive{\text{S}^{5}}
\def\AdSS{\text{AdS}_{5}\times\text{S}^{5}}
\def\truN{\mathop{\text{tr}_{\text{U}(N)}}\nolimits}
\def\truK{\mathop{\text{tr}_{\text{U}(K)}}\nolimits}
\def\Str{\mathop{\text{Str}}\nolimits}
\def\vy{{\vec y\,}}\def\ve{{\vec\epsilon\,}}
\theoremstyle{plain}

\theoremstyle{definition}
\theoremstyle{remark}

\def\plb#1#2#3{{\it Phys.\ Lett.\ }{\bf B #1} (#2) #3}
\def\npb#1#2#3{{\it Nucl.\ Phys.\ }{\bf B #1} (#2) #3}

\def\jhep#1#2#3{{\it J. High Energy Phys.\ }{\bf #1} (#2) #3}
\def\prd#1#2#3{{\it Phys.\ Rev.\ }{\bf D #1} (#2) #3}

\def\atmp#1#2#3{{\it Adv.\ Theor.\ Math.\ Phys.\ }{\bf #1} (#2) #3}

\def\pr#1#2#3{{\it Phys.\ Rep.\ }{\bf #1} (#2) #3}

\def\imath#1#2#3{{\it Invent math }{\bf #1} (#2) #3}
\def\jgeomphys#1#2#3{{\it J.\ Geom.\ Phys.\ }{\bf #1} (#2) #3}
\begin{document}
%
%
{\pagestyle{empty}
\parskip 0in
\
\vfill
\begin{center}
{\LARGE Emergent Space and the Example of $\AdSS$}


\vspace{0.4in}

Frank F{\scshape errari}
\\
\medskip
{\it Service de Physique Th\'eorique et Math\'ematique\\
Universit\'e Libre de Bruxelles and International Solvay Institutes\\
Campus de la Plaine, CP 231, B-1050 Bruxelles, Belgique}
\smallskip
{\tt frank.ferrari@ulb.ac.be}
\end{center}
\vfill\noindent

We explain how to build field theoretic observables from which the geometrical properties of a dual holographic formulation can be read off straightforwardly. In some cases this construction yields explicit and calculable models of emergent space. We illustrate the idea on the type IIB background generated by $N$ D3-branes in the near horizon limit, for which a full derivation from first principles can be presented. The six transverse dimensions emerge at large $N$  and we find the full $\AdSS$ metric and self-dual Ramond-Ramond field strength on the resulting ten dimensional space-time, with the correct radii and quantization law. We briefly discuss possible applications and generalizations.

\vfill

\medskip
%
\begin{flushleft}
\today
\end{flushleft}

\newpage\pagestyle{plain}
\baselineskip 16pt
\setcounter{footnote}{0}

}

\section{\label{s1} Introduction}

The correct starting point for a consistent theory of quantum gravity remains, to the present day, elusive. The most straightforward approaches, which are suggested by the classical description of gravity in terms of a metric tensor, are plagued by well-known and possibly insurmountable
difficulties, at least in space-time dimension four or higher. The problem is much deeper than perturbative non-renormalizability, or even our lack of understanding of the space of metrics on a given manifold. It is related to
the breakdown of our most cherished tool of standard local quantum field theory, the renormalization group. This seems to be an inescapable consequence of fundamental properties of quantum gravity, like background independence, general covariance and the lack of local observables.

These deep conceptual issues suggest that gravity may be of an entirely different nature than the other known forces which are described by local quantum fields. A simple and beautiful idea, that dates back to Sakharov in the 60s \cite{Sak}, is to consider that gravity could be an emergent phenomenon. In other words, the gravitational force and its geometric description within Einstein's general relativity would correspond to an approximate description, valid in some regime, of an underlying pre-geometric microscopic model whose formulation does not refer to gravity. In view of the plethora of emergent phenomena in physics, and consistently with the well-known formal links between general relativity and thermodynamics, this point of view seems very natural.

The simplest scenarios of emergent gravity are ruled out by a famous no-go theorem by Weinberg and Witten \cite{WW}, which makes it very difficult to build simple models consistent with general covariance and local Lorentz invariance (for a recent review, see e.g.\ \cite{emergereview}).
An attractive way out of the Weinberg-Witten theorem is to assume that \emph{a theory of emergent gravity must also be a 
theory of emergent space.} In other words, the very notion of space should be approximate and emerge alongside with geometric properties like the metric and the other physical fields propa\-gating on it. A nice discussion of this idea can be found for example in 
\cite{Seibergemergent}. This point of view is realized in the context of the AdS/CFT correspondence \cite{MaldaAdS} and could very well be the correct starting point for string theory and quantum gravity in general. The main difficulty one encounters by trying to follow this path is to build \emph{calculable} models. In the literature on AdS/CFT, this shows through the huge imbalance on the way the correspondence is used, with most works focusing on the study of strongly coupled field theory from classical gravity and comparatively very few trying to understand quantum gravity from field theory \cite{bererev}. The origin of this bias is purely technical: classical gravity is much easier to deal with than strongly coupled large $N$ field theories.

Our aim in this paper is to present a simple calculation, from first principles, which shows clearly how \emph{the full type IIB $\AdSS$ background, including the correctly quantized Ramond-Ramond self-dual five-form field strength,} emerges from a purely field-theoretical, pre-geometric microscopic model. To our knowledge, and in spite of many impressive results in the literature (see e.g.\ \cite{bererev} for a review and \cite{instbrit1,akh,rey} for a few examples), this has never been obtained by any other method. A basic difficulty is that typical field theory calculations yield expansions in the coupling constants, from which it is highly non-trivial to find hints about a possible geometrical interpretation. \emph{We shall bypass this difficulty by building particular observables in the field theory from which all the supergravity fields can be read off straightforwardly.}

The method can be applied to a variety of cases and actually suggests a general strategy to build models of emergent space without referring explicitly to a particular string theory framework \cite{toappearc}. Presently, we shall limit our explicit calculations to the simplest set-up of the near-horizon D3-brane background. In this case, our work is clearly related to, and may provide a justification of, remarkable previous works in the literature \cite{instbrit1,akh}. In particular, it would be interesting to make the link with the striking instanton calculations made in \cite{instbrit1} explicit, but we shall not attempt to do that in the following.
We have also studied other geometries, including the near-horizon D4-brane background and non-supersymmetric cases. These additional results will appear in forthcoming publications \cite{toappeara,toappearb,toappearc}.

Motivated by the fact that many different topics will enter into our discussion (AdS/CFT, instantons, non-abelian D-brane actions, vector models etc...) we have tried to be as pedagogical as possible. We start by presenting the general framework in section \ref{s2}, explaining in particular the slight extension of the standard AdS/CFT set-up that we use. In section \ref{s3}, we provide full details on the field theoretic (open string) aspects of the D$3$/D$(-1)$ system on which we focus. To solve the resulting model, we have to compute a seemingly non-trivial sum over planar diagrams.
We explain in section \ref{s4} that these diagrams divide in two classes.  
The diagrams in the first class can always be summed up explicitly.
The bulk holographic space dimensions emerge from this sum. The sum over the second class of diagrams is in general intractable, but a simple argument shows that, at least when conformal invariance is present, it is actually trivial, yielding an exactly solvable model. 
In section \ref{s5}, we explain how the metric and the other supergravity fields propagating on the emergent space can then be straightforwardly derived. Sections \ref{s4} and \ref{s5} together provide an explicit mapping between any state of the $\nn=4$ super Yang-Mills theory and a type IIB background.
The detailed calculation of the background associated with the conformal vacuum is presented in section \ref{s6}. Some possible applications, extensions and implications of our work are finally discussed in section \ref{s7}.

\section{\label{s2} The framework}
\begin{figure}
\centerline{\includegraphics[width=5in]{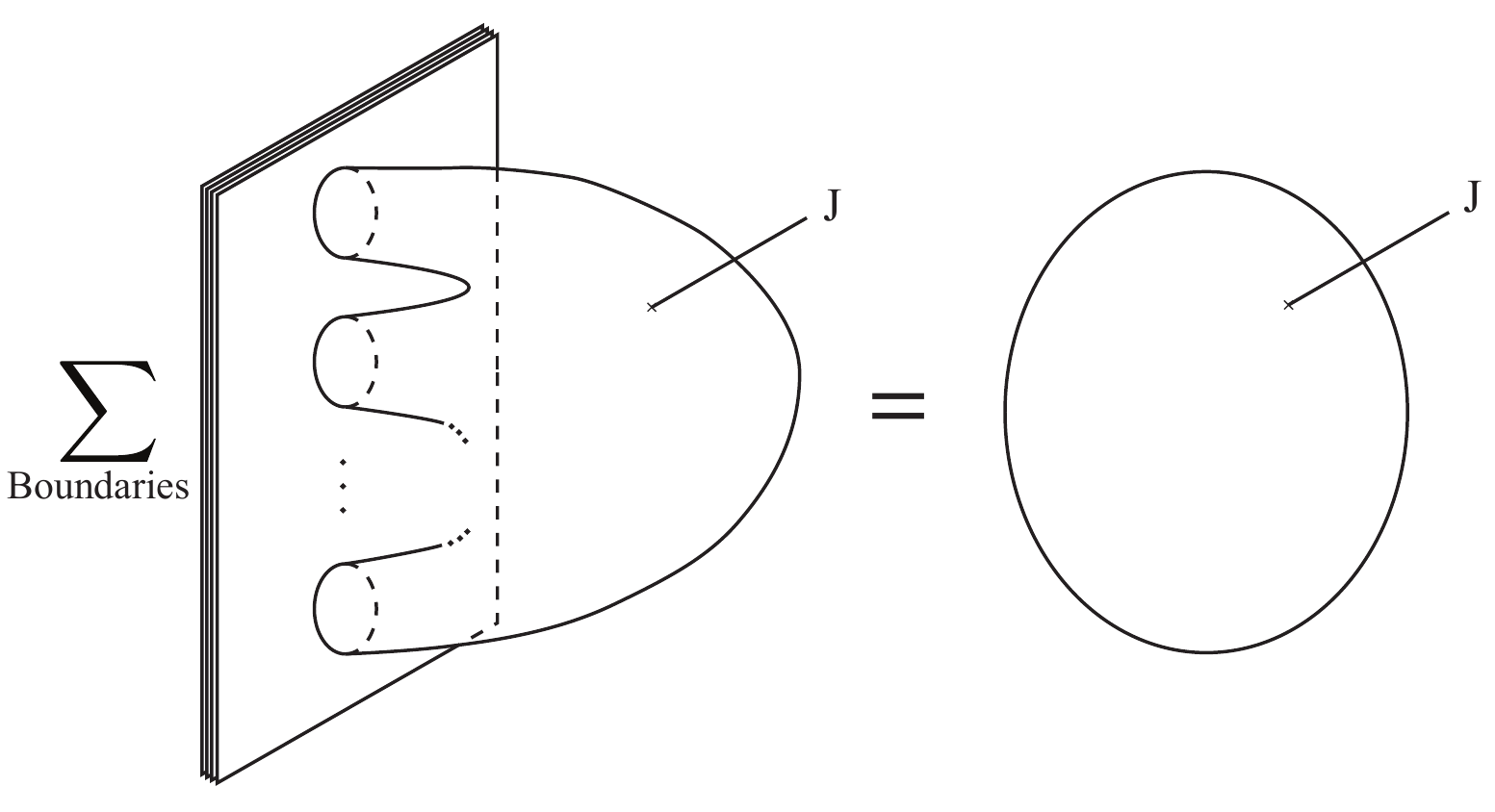}}
\caption{On the left, the world-sheets describing the leading large $N$ interaction between a bulk closed string mode with wave function $J$ and a stack of $N$ background branes. The number of boundaries on the background branes world volume can be arbitrary, corresponding to a sum over loops in the microscopic gauge theoretic path integral \eqref{pathint0}. This sum is replaced on the right by a unique closed string sphere diagram in a non-trivial geometrical background.
 \label{fig1}}
\end{figure}

An example of the usual open/closed string duality is depicted on figure \ref{fig1}. A closed string mode, for example a graviton with wave function $J$, scatter off a stack of $N$ background D-branes. The interaction between the closed string and the branes is described by string diagrams proportional to $(\gs N)^{B}\gs^{2h}$, where $\gs$ is the string coupling constant, $B$ the number of world-sheet boundaries on the branes and $h$ the number of world-sheet handles. In the usual large $N$ limit at fixed 't~Hooft's coupling
\be\label{lambdadef} \lambda = 4\pi N\gs\, , 
\ee
the diagrams on the left of the figure, with $h=0$, dominate. At low energy, these diagrams correspond to the $L=B-1$ loops planar diagrams of the $\uN$ gauge theory living on the brane. These diagrams compute a large $N$ path integral of the form 
\be\label{pathint0}\int\!\d^{N^{2}}\!\!\mu_{\text b}\, e^{-S_{\text b}+\int\! J\mathscr O}\, ,\ee
where $\d^{N^{2}}\!\mu_{\text b}$ is the integration measure over the $\sim N^{2}$ degrees of freedom in the adjoint of $\text{U}(N)$ living on the background branes world-volume, $S_{\text b}$ the associated world-volume action and $\mathscr O$ the local gauge invariant operator on the branes to which the closed string mode couples. The open/closed string duality claims that the sum of diagrams on the left-hand side of figure \ref{fig1} should be equivalent to the unique closed string diagram represented on the right-hand side of the figure, in which the background branes have been replaced by a non-trivial emergent type IIB background. This diagram computes a gravitational effective action $S_{\text{grav}}(J)$ describing the propagation of the closed string mode in the non-trivial background.
This propagation is described by the supergravity field equations
including the possible $\apr$ corrections and the equality illustrated in figure \ref{fig1} is equivalent to
\be\label{pathint0b}\int\!\d^{N^{2}}\!\!\mu_{\text b}\, e^{-S_{\text b}+\int\! J\mathscr O} = e^{-S_{\text{grav}}(J)}\, ,\ee
which is the standard AdS/CFT correspondence. The above equation is also supposed to be valid at finite $N$, as long as string quantum corrections are taken into account in $S_{\text{grav}}$.

Equation \eqref{pathint0b} is clearly a spectacular example of emergent geometry. However, even if we could compute the field theoretic generating function on the left-hand side, it would be highly non-trivial to guess that it admits the geometrical interpretation in higher dimensional space given by the right-hand side. One would like to have a version of the correspondence where the emergent space and its geometrical properties appear more explicitly and are thus more amenable to study directly from field theory.

\begin{figure}
\centerline{\includegraphics[width=5in]{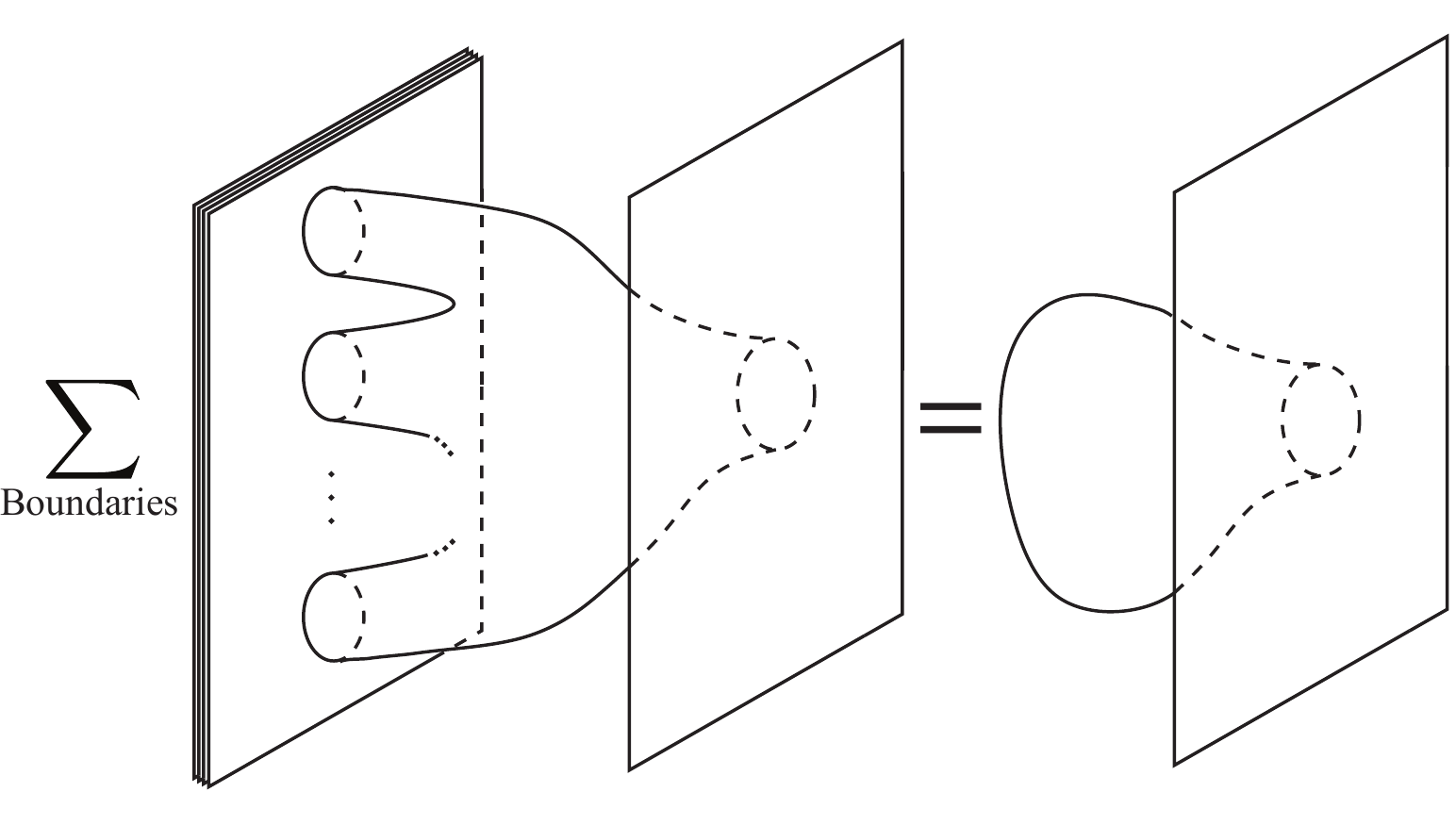}}
\caption{On the left, the world-sheets describing the leading large $N$ interaction between $K$ probe branes and a stack of $N$ background branes. The number of boundaries on the background branes world volume can be arbitrary, corresponding to a sum over loops in the microscopic gauge theoretic path integral \eqref{pathint1}. This sum is replaced on the right by a unique open string disk diagram in a non-trivial geometrical background. \label{fig2}}
\end{figure}

For this purpose, consider the slightly modified version of the correspondence depicted in figure \ref{fig2}.
A stack of a large number $N$ of background D-branes and a fixed number $K$ of probe branes interact via string diagrams having boundaries on both type of branes. A diagram of this type is proportional to
$(\gs N)^{B}(\gs K)^{b}\gs^{2 h}$, where $B$ and $b$ are the numbers of world-sheet boundaries on the background and probe branes respectively and $h$ the number of world-sheet handles. In the 't~Hooft limit, the interaction is dominated by the sum over $h=0$, $b=1$ diagrams depicted 
on the left-hand side of figure \ref{fig2}. The sum over these diagrams computes a large $N$ path integral of the form
\be\label{pathint1} \int\!\d^{N^{2}}\!\!\mu_{\text{b}}\d^{N}\!\mu_{\text{p}}\,
e^{-S_{\text{b}}-S_{\text{p}}}\, ,\ee
where $\d^{N^{2}}\!\!\!\mu_{\text b}$ and $\d^{N}\!\mu_{\text p}$ are the integration measures over the degrees of freedom living on the background ($\sim N^{2}$ fields in the adjoint of $\text{U}(N)$) and probe ($\sim N$ fields in the fundamental and anti-fundamental of $\text{U}(N)$) world-volumes and $S_{\text b}$ and $S_{\text p}$ are the associated world-volume actions. This sum
should be equivalent to the unique diagram represented on the right-hand side of figure \ref{fig2}, in which the background branes have been replaced by a non-trivial emergent type IIB background. 
This diagram, with possible insertions of open string vertex operators on its boundary, computes an effective action $S_{\text{eff}}$ for the probe/probe strings degrees of freedom in the adjoint of $\text{U}(K)$ moving in the non-trivial background. 
If we denote these degrees of freedom by $Z$ and $\Psi$ for the bosons and fermions respectively,  
then the equality illustrated in figure \ref{fig2} is equivalent to 
\be\label{pathint2} \int\!\d^{N^{2}}\!\!\mu_{\text{b}}\d^{N}\!\mu_{\text{p}}\,
e^{-S_{\text{b}}-S_{\text{p}}}=
\int\!\d Z\d\Psi\, e^{-S_{\text{eff}}(Z,\Psi)}\, .\ee
Any number of insertions of $\text{U}(K)$ invariant operators depending on $Z$ and $\Psi$ could be added in \eqref{pathint2}.

Equations \eqref{pathint0b} and \eqref{pathint2} are conceptually very similar but the emergent geometry and the fields propagating on it can be read off much more explicitly from $S_{\text{eff}}(Z,\Psi)$ than from the generating function $S_{\text{grav}}(J)$. The effective action $S_{\text{eff}}(Z,\Psi)$ can itself be expressed in terms of specific gauge invariant correlators in the background brane theory, as explained in details in section \ref{s5}.
 
An important point to make is that the total number of variables $(Z,\Psi)$, which are $\text{U}(N)$ singlets, is independent of $N$. On the other hand, the brane effective action $S_{\text{eff}}$ is always proportional to $1/\gs\sim N$ in the 't~Hooft limit. This implies that the fluctuations of $Z$ and $\Psi$ are suppressed at large $N$. 
The bosonic variables $Z$ can then be interpreted as classical matrix coordinates for the probe branes: a classical space has emerged.

\section{The case of the D3/D(-1) system}
\label{s3}
\begin{figure}
\centerline{\includegraphics[width=5in]{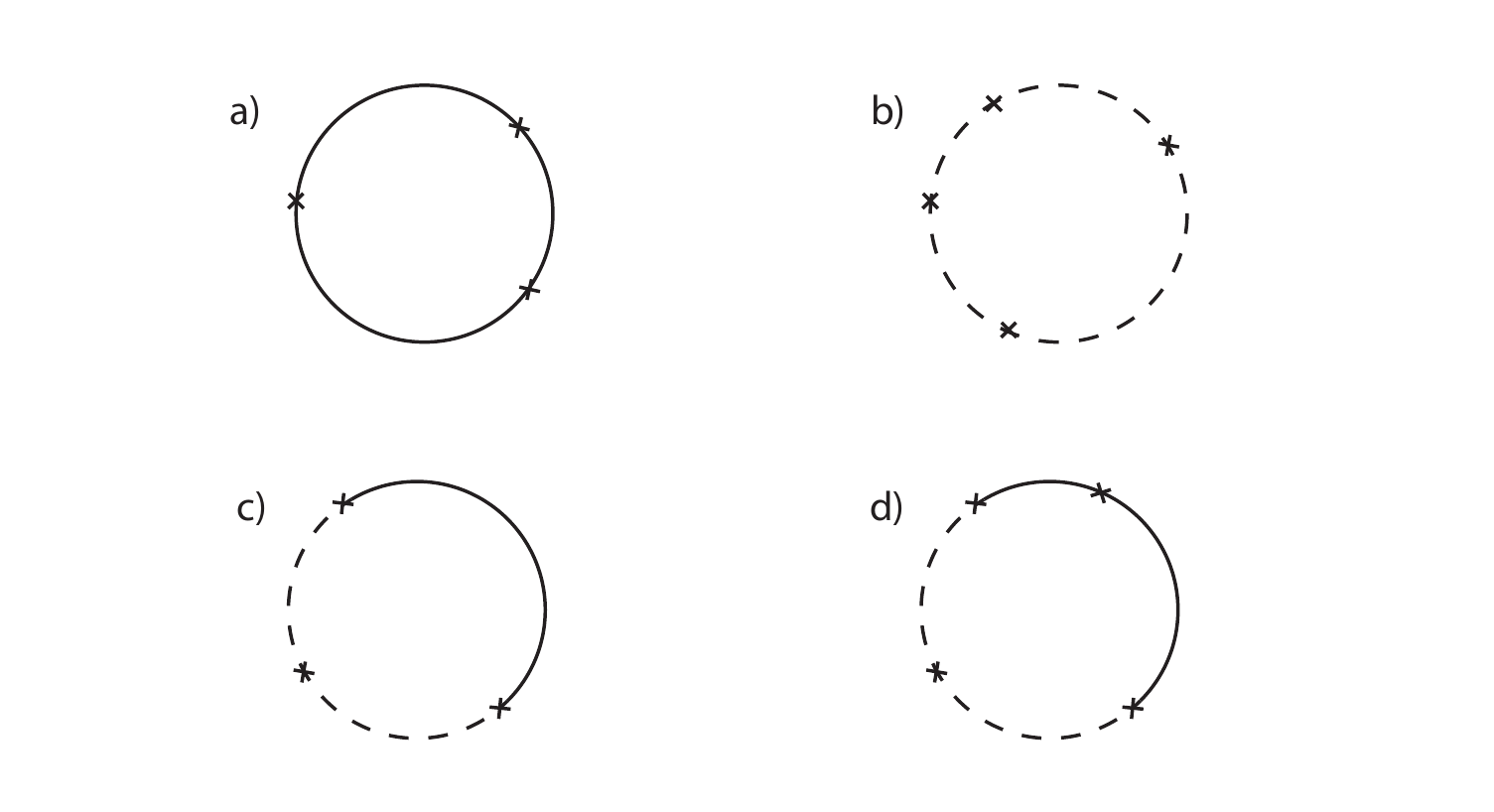}}
\caption{Disk diagrams contributing to the action 
$S_{\text b}+S_{\text p}$ \cite{couplings}. The plain lines represent boundaries on the background branes, the dashed lines boundaries on the probe branes. a) Diagrams with the world-sheet boundary on the background D3 branes compute the $\nn=4$ super Yang-Mills action \eqref{Sb}. A cubic term, e.g.\ a Yukawa coupling, is explicitly represented; b) Diagrams with the world-sheet boundary on the probe D-instantons compute the action \eqref{Sp1}. The quartic $[X,\phi]^{2}$ term in \eqref{Sp1} is explicitly represented; c) and d) Diagrams with mixed boundary conditions, without and with vertex operator insertions on the boundary situated on the background branes, compute the action \eqref{Sp31t} and the couplings to the $\mathcal N=4$ local fields respectively. In c), a cubic term in \eqref{Sp31t} is represented, for example the $\tilde q\Lambda\chi$ coupling. In d), the quartic $\tilde q\phi\varphi q$ coupling in \eqref{Sp31} is represented.\label{fig3}}
\end{figure}

Let us now focus on the particular case of background D3-branes and probe D-instantons in type IIB string theory and write down explicitly the path integral in \eqref{pathint1}. The total action $S_{\text b}+S_{\text p}$ can be obtained by considering the appropriate low energy limit of various open string disk diagrams \cite{couplings}, some of which are depicted in figure \ref{fig3}. 

The diagrams of type a), with boundary on the background branes, compute the background brane action $S_{\text b}$. The result is of course well-known: one obtains the maximally supersymmetric $\mathcal N=4$ gauge theory with gauge group $\uN$, global symmetry group (in the euclidean) $\text{Spin}(4)\times \text{SU}(4)$, elementary fields
\be\label{fieldsN4} 
(a_{\mu},\varphi_{A},\lambda_{\alpha a},\bar\lambda^{\dot\alpha a})
\ee
in the adjoint of $\uN$ and action
\begin{multline}\label{Sb} S_{\text b}=\frac{1}{4\pi\gs}\truN\int\!\d^{4}x\,\Bigl\{
\frac{1}{2}F_{\mu\nu}F_{\mu\nu} + \nabla_{\mu}\varphi_{A}\nabla_{\mu}
\varphi_{A} - \frac{1}{2}\bigl[\varphi_{A},\varphi_{B}\bigr]
\bigl[\varphi_{A},\varphi_{B}\bigr]\\
+2 i \lambda^{\alpha}_{\ a}\sigma_{\mu\alpha\dot\alpha}\nabla_{\mu}\bar
\lambda^{\dot\alpha a} 
- \lambda^{\alpha}_{\ a}\bar\Sigma_{A}^{ab}
\bigl[\varphi_{A},\lambda_{\alpha b}\bigr]
- \bar\lambda_{\dot\alpha}^{\ a}\Sigma_{Aab}
\bigl[\varphi_{A},\bar\lambda^{\dot\alpha b}\bigr]\Bigr\}\, .
\end{multline}
This action is invariant under sixteen supercharges $Q^{\alpha a}$ and $\bar Q_{\dot\alpha a}$.
Our conventions for the meaning of the various indices and the associated transformation laws of the fields are explained in details in the table \ref{indices} of the appendix at the end of the paper. The $\sigma_{\mu\alpha\dot\alpha}$, $\Sigma_{Aab}$ and $\bar\Sigma_{A}^{ab}$ are the Clebsch-Gordan coefficients coupling the vector and spinors of $\text{SO}(4)$ and $\text{SO}(6)$ respectively. Explicit formulas for these coefficients as well as several identities that are used later in the text can also be found in the appendix. 

The diagrams of type b), with boundary on the probe branes, compute the part $S_{\text p,-1/-1}$ of the probe brane action associated with the D$(-1)$/D$(-1)$ strings in the adjoint of $\text{U}(K)$. Up to an important subtlety that we discuss next, this action can actually be obtained straightforwardly from the $\mathcal N=4$ super Yang-Mills action \eqref{Sb} associated with the diagrams of type a). Indeed, in strict parallel with the fact that \eqref{Sb} can be viewed as the dimensional reduction of the ten dimensional super Yang-Mills theory down to four dimensions, $S_{\text p,-1/-1}$ is related to the dimensional reduction of the same theory but down to zero dimension. Of course this is also equivalent to dimensionally reduce \eqref{Sb} down to zero dimension. The fields
in $S_{\text p,-1/-1}$ thus have exactly the same structure as in
\eqref{fieldsN4} and will be denoted by
\be\label{modulimatrix} 
(A_{\mu},\phi_{A},\Lambda_{\alpha a},\bar\Lambda^{\dot\alpha a})\, .
\ee
Actually, the ``fields'' \eqref{modulimatrix} are better called moduli, since they do not depend on any space-time point.
The action can be obtained from $\eqref{Sb}$ by substituting $a_{\mu}\rightarrow A_{\mu}$, $\varphi_{A}\rightarrow\phi_{A}$, $\lambda_{\alpha a}\rightarrow\Lambda_{\alpha a}$,
$\bar\lambda^{\dot\alpha a}\rightarrow\bar\Lambda^{\dot\alpha a}$, changing the gauge group from $\text{U}(N)$ to $\text{U}(K)$ and doing the dimensional reduction. This yields
\begin{multline}\label{Sp1bis} \tilde S_{\text p,-1/-1}=\frac{\pi\ls^{4}}{\gs}\truK \Bigl\{
-\frac{1}{2}\bigl[A_{\mu},A_{\nu}\bigr]\bigl[A_{\mu},A_{\nu}\bigr] -
\bigl[A_{\mu},\phi_{A}\bigr]\bigl[A_{\mu},\phi_{A}\bigr]
 - \frac{1}{2}\bigl[\phi_{A},\phi_{B}\bigr]
\bigl[\phi_{A},\phi_{B}\bigr]\\
-2  \Lambda^{\alpha}_{\ a}\sigma_{\mu\alpha\dot\alpha}
\bigl[A_{\mu},\bar\Lambda^{\dot\alpha a} \bigr]
- \Lambda^{\alpha}_{\ a}\bar\Sigma_{A}^{ab}
\bigl[\phi_{A},\Lambda_{\alpha b}\bigr]
- \bar\Lambda_{\dot\alpha}^{\ a}\Sigma_{Aab}
\bigl[\phi_{A},\bar\Lambda^{\dot\alpha b}\bigr]\Bigr\}\, .
\end{multline}
This action is invariant under the same sixteen supercharges $Q^{\alpha a}$ and $\bar Q_{\dot\alpha a}$ as \eqref{Sb}. It also has the full $\text{Spin}(10)$ global symmetry group, of which only the 
$\text{Spin}(4)\times\text{SU}(4)$ subgroup is made manifest in our notations; this is the most natural point of view, since $\text{Spin}(10)$ is broken by the other pieces in the action $S_{\text p}$, coming from the diagrams a), c) and d) in figure \ref{fig3}. The only non-trivial input from string theory is the overall numerical coefficient in front of \eqref{Sp1bis}.

The action \eqref{Sp1bis}, however, is not exactly what we need. We have to be careful about the precise implementation of the low-energy limit associated with the standard near-horizon limit in the background of the D3-branes. The coordinates transverse to the D3 branes are associated with the bosonic moduli $\ls^{2}\phi_{A}$ (taking into account that the dimension of $\phi_{A}$ is an inverse length), and the near horizon limit must be taken by keeping 
$\ls^{2}\phi_{A}/\ls^{2}=\phi_{A}$ fixed \cite{MaldaAdS}. On the other hand, the coordinates parallel to the D3-branes are associated with the moduli
\be\label{Xmudef}
X_{\mu}=\ls^{2}A_{\mu}\ee
and these coordinates must also be kept fixed in the limit. Similar scalings must be imposed on the fermions, which are the supersymmetric partners of the bosonic moduli. There are two completely equivalent possibilities at this stage, depending on which supercharges one wishes to preserve for the D$3$/D$(-1)$ system (or, equivalently, if one wishes to consider D-instantons or anti D-instantons). If we pick, for example, the eight supercharges $Q^{\alpha a}$, then the superpartner of $\phi_{A}$ is $\Lambda_{\alpha a}$, which is thus kept fixed, and the superpartner of $X_{\mu}$ is 
\be\label{barpsidef}
\bar\psi^{\dot\alpha a}=\ls^{2}\bar\Lambda^{\dot\alpha a}\, ,\ee
which is also kept fixed in the limit. Taking $\ls\rightarrow 0$ in \eqref{Sp1bis} then yields
\begin{multline}\label{Sp1tierce} \tilde S_{\text p,-1/-1}'=\frac{\pi}{\gs}\truK \Bigl\{
-\frac{1}{2\ls^{4}}\bigl[X_{\mu},X_{\nu}\bigr]\bigl[X_{\mu},X_{\nu}\bigr] -
\bigl[X_{\mu},\phi_{A}\bigr]\bigl[X_{\mu},\phi_{A}\bigr]\\
-2  \Lambda^{\alpha}_{\ a}\sigma_{\mu\alpha\dot\alpha}
\bigl[X_{\mu},\bar\psi^{\dot\alpha a} \bigr]
- \bar\psi_{\dot\alpha}^{\ a}\Sigma_{Aab}
\bigl[\phi_{A},\bar\psi^{\dot\alpha b}\bigr]\Bigr\}\, .
\end{multline}
To deal with the singular term in $1/\ls^{4}$, it is convenient to introduce an auxiliary field $D_{\mu\nu}$ in the adjoint of $\text{U}(K)$ and make the replacement
\be\label{auxiliary}-\frac{1}{2\ls^{4}}\truK \bigl[X_{\mu},X_{\nu}\bigr]\bigl[X_{\mu},X_{\nu}\bigr]\rightarrow \truK\Bigl\{ -\ls^{4}D_{\mu\nu}D_{\mu\nu} + 2i
D_{\mu\nu}\bigl[X_{\mu},X_{\nu}\bigr]\Bigr\}\, .\ee
One can impose a duality constraint on $D$,
\be\label{Dself} D_{\mu\nu}=\pm\frac{1}{2}\epsilon_{\mu\nu\rho\sigma}
D_{\rho\sigma}\ee
and use the identity
\be\label{idepsilon} \epsilon_{\mu\nu\rho\sigma}\tr [X_{\mu},X_{\nu}]
[X_{\rho},X_{\sigma}] = 0\ee
to show that the correct original quartic term is obtained from the right-hand side of \eqref{auxiliary} by integrating out $D_{\mu\nu}$. With our choice of conserved supercharges $Q^{\alpha a}$ it is convenient to choose the plus sign in \eqref{Dself}. When $\ls\rightarrow 0$, the quadradic term in \eqref{auxiliary} vanishes and we finally get the correct low energy limit of \eqref{Sp1bis},
\begin{multline}\label{Sp1} S_{\text p,-1/-1}=\frac{\pi}{\gs}\truK \Bigl\{
2iD_{\mu\nu}\bigl[X_{\mu},X_{\nu}\bigr] -
\bigl[X_{\mu},\phi_{A}\bigr]\bigl[X_{\mu},\phi_{A}\bigr]\\
-2  \Lambda^{\alpha}_{\ a}\sigma_{\mu\alpha\dot\alpha}
\bigl[X_{\mu},\bar\psi^{\dot\alpha a} \bigr]
- \bar\psi_{\dot\alpha}^{\ a}\Sigma_{Aab}
\bigl[\phi_{A},\bar\psi^{\dot\alpha b}\bigr]\Bigr\}\, .
\end{multline}
The reader familiar with supersymmetry will recognize in the triplet $(\phi_{A},\Lambda_{\alpha a},D_{\mu\nu})$ the off-shell vector multiplet of six dimensional $\nn=1$ super Yang-Mills (or equivalently the off-shell $\nn=2$ vector multiplet in four dimensions), whereas $(X_{\mu},\bar\psi^{\dot\alpha a})$ forms an adjoint hypermultiplet. Supersymmetry transformation laws are indicated for completeness in appendix \ref{apB}.

The diagrams of types c) and d) come with insertions of boundary changing vertex operators. These operators correspond to hypermultiplet degrees of freedom 
\be\label{modulivector} (q_{\alpha},\chi^{a},\tilde q^{\alpha},\tilde\chi^{a})\ee
associated with the D$3$/D$(-1)$ strings in the bi- (or anti bi-) fundamental representation of the gauge group $\text{U}(N)\times\text{U}(K)$. Using, as previously, the auxiliary field $D_{\mu\nu}$ to deal with the singular terms occurring in the limit $\ls\rightarrow 0$, 
the diagrams c) generate the action
\begin{multline}\label{Sp31t} \tilde S_{\text p,-1/3}= \frac{i}{2}
\tilde q^{\alpha}D_{\mu\nu}\sigma_{\mu\nu\alpha}^{\hphantom{\mu\nu\alpha}\beta}q_{\beta}+\frac{1}{2}\tilde q^{\alpha}\phi_{A}\phi_{A}q_{\alpha}-\frac{1}{2}
\tilde\chi^{a}\Sigma_{Aab}\phi_{A}\chi^{b}\\+ \frac{1}{\sqrt{2}}
\tilde q^{\alpha}\Lambda_{\alpha a}\chi^{a} + \frac{1}{\sqrt{2}}
\tilde\chi^{a}\Lambda^{\alpha}_{\ a}q_{\alpha}\, . \end{multline}
This action is invariant under the supersymmetry transformations indicated in appendix \ref{apB}.

The result of the above discussion, whose aim was mainly to be pedagogical and to present the reasoning in line with the AdS/CFT framework discussed in section \ref{s2}, is of course well known \cite{Dbranesandinst,instrev}. The sum of \eqref{Sp1} and \eqref{Sp31t} yields the action for the ADHM instanton moduli of the $\mathcal N=4$ Yang-Mills theory living on the D$3$ branes, the ADHM constraints being implemented by the moduli $D_{\mu\nu}$ and $\Lambda_{\alpha a}$ which, since they appear only linearly, play the r\^ole of Lagrange multipliers.

There are, however, other important terms in the probe action. These terms are associated with the diagrams of type d), describing the couplings between the moduli and the local fields of the $\nn=4$ theory. These couplings have been studied in details in \cite{couplings}. Their effect is to transform the action \eqref{Sp31t} by shifting the moduli in the vector multiplet $(\phi_{A},\Lambda_{\alpha a},D_{\mu\nu})$ by corresponding local fields $(\hat\varphi_{A},\hat\lambda_{\alpha a},\hat F_{\mu\nu}^{+})$ in the $\nn=4$ theory. Writing explicitly all the gauge indices for clarity, this yields the following action,
\begin{multline}\label{Sp31} S_{\text p,-1/3}= \frac{i}{2}
\tilde q^{\alpha fi}\bigl(D_{\mu\nu i}^{\ \ \ j}\delta_{f}^{f'} - \delta_{i}^{j}\hat F_{\mu\nu f}^{+\ f'}\bigr)\sigma_{\mu\nu\alpha}^{\hphantom{\mu\nu\alpha}\beta}q_{\beta f'j}\\
+\frac{1}{2}\tilde q^{\alpha fi}
\bigl(\phi_{Ai}^{\ \ j}\delta_{f}^{f'} - \delta_{i}^{j}\hat\varphi_{Af}^{\ \ f'}\bigr)  \bigl(\phi_{Aj}^{\ \ k}\delta_{f'}^{f''} - \delta_{j}^{k}\hat\varphi_{Af'}^{\ \ f''}\bigr)
q_{\alpha f''k}
-\frac{1}{2}
\tilde\chi^{afi}\Sigma_{Aab}\bigl(\phi_{Ai}^{\ \ j}\delta_{f}^{f'}-\delta_{i}^{j}\hat\varphi_{Af}^{\ \ f'}\bigr)\chi^{b}_{\ f'j}\\
+ \frac{1}{\sqrt{2}}
\tilde q^{\alpha fi}\bigl(\Lambda_{\alpha a i}^{\ \ \ j}\delta_{f}^{f'}-\delta_{i}^{j}\hat\lambda_{\alpha a f}^{\ \ \ f'}\bigr)
\chi^{a}_{\ f'j}
+ \frac{1}{\sqrt{2}}
\tilde\chi^{afi}\bigl(\Lambda^{\alpha\ j}_{\ ai}\delta_{f}^{f'} -
\hat\lambda^{\alpha\ f'}_{\ af}\delta_{i}^{j}\bigr)
q_{\alpha f'j}\, . \end{multline}
The fields $(\hat\varphi_{A},\hat\lambda_{\alpha a},\hat F_{\mu\nu}^{+})$ can be expanded in powers of the moduli 
\be\label{hatpsidef}\hat{\bar\psi}^{\dot\alpha a} = \frac{1}{K}\truK
\bar\psi^{\dot\alpha a}\ee
as
\begin{align}\nonumber
\hat\varphi_{A} & = \varphi_{A}(\hat x) + \sum_{n\geq 1}\mathscr 
O_{A\dot\alpha_{1}a_{1}\cdots\dot\alpha_{n}a_{n}}(\hat x)
\hat{\bar\psi}^{\dot\alpha_{1}a_{1}}\cdots\hat{\bar\psi}^{\dot\alpha_{n}a_{n}}\\\nonumber
\hat\lambda_{\alpha a} & = \lambda_{\alpha a}(\hat x) + \sum_{n\geq 1}\mathscr 
O_{\alpha a\dot\alpha_{1}a_{1}\cdots\dot\alpha_{n}a_{n}}(\hat x)
\hat{\bar\psi}^{\dot\alpha_{1}a_{1}}\cdots\hat{\bar\psi}^{\dot\alpha_{n}a_{n}}\\\label{locexp}
\hat F_{\mu\nu}^{+} & = F_{\mu\nu}^{+}(\hat x) + \sum_{n\geq 1}\mathscr 
O_{\mu\nu\dot\alpha_{1}a_{1}\cdots\dot\alpha_{n}a_{n}}(\hat x)
\hat{\bar\psi}^{\dot\alpha_{1}a_{1}}\cdots\hat{\bar\psi}^{\dot\alpha_{n}a_{n}}\, ,
\end{align}
where $\varphi_{A}$ and $\lambda_{\alpha a}$ are the local fields in the action \eqref{Sb} and $F_{\mu\nu}^{+}$ the self-dual part of the field-strength. Of course, the sums in \eqref{locexp} have a finite number of terms. Explicit formulas for the various local operators $\mathscr O$ can be found \cite{couplings}, but these details will not be needed for our purposes. The only feature that we shall use (in particular in the next section) is that all the $\nn=4$ local operators that appear in \eqref{locexp} are evaluated at the same point,
\be\label{hatxdef} \hat x_{\mu}=\frac{1}{K}\truK X_{\mu}\, ,\ee
which corresponds to the center-of-mass position of the $K$ D-instantons.

We have thus obtained a detailed definition of the path integral \eqref{pathint1}. The action $S_{\text b}$ is given by \eqref{Sb}, the action $S_{\text p}$ is the sum of the actions 
\eqref{Sp1} and \eqref{Sp31},
\be\label{Sp} S_{\text p}= S_{\text p,-1/-1}+ S_{\text p,-1/3}\ee
and the integration measure corresponds to the standard flat measure on the fields and moduli $(a_{\mu},\varphi_{A},\lambda,\bar\lambda,X_{\mu},\bar\psi,\phi_{A},\Lambda,D_{\mu\nu},q,\chi,\tilde q,\tilde\chi)$ discussed above, together with ghosts associated with the $\text{U}(N)$ gauge-fixing procedure whose details will be of no concern for us.

\section{\label{s4} Summing up diagrams}

The computation of \eqref{pathint1} goes in two steps
which, at least in the particular example we are studying, turn out to be technically very simple but conceptually quite interesting. 
Combined with the discussion of section \ref{s5}, we are going to construct a precise correspondence mapping any state in the theory on the background branes to a ten-dimensional geometry.

\begin{figure}
\centerline{\includegraphics[width=5in]{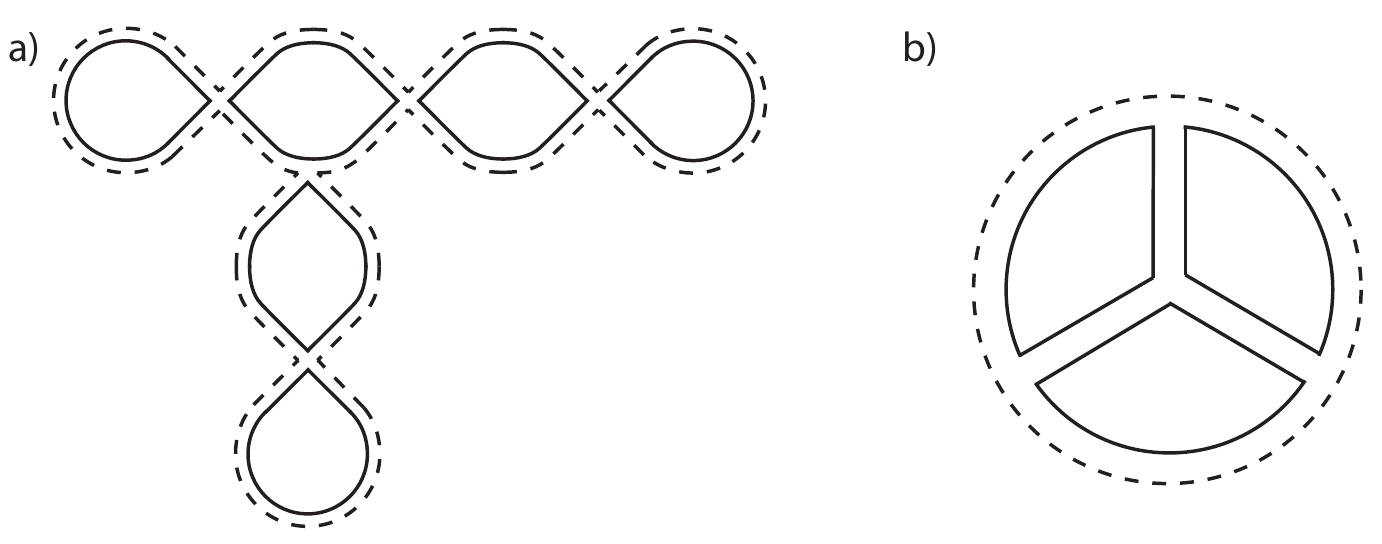}}
\caption{Typical multi-loop field theory diagrams associated with the path integral \eqref{pathintFT} or equivalently with the string diagrams of the left of figure \ref{fig2}. The dashed lines correspond to \text{U}(K) indices (world-sheet boundary on the probe branes) and the plain lines to $\text{U}(N)$ indices (world-sheet boundary on the background branes) as in figure \ref{fig3}. a) A tree-like bubble diagram typical of vector models; b) A typical matrix model planar diagram.\label{fig4}}
\end{figure}
\subsection{\label{sub41} Vector model diagrams}

In the previous section, we have introduced the actions \eqref{Sp1} and \eqref{Sp31} from the point of view of open string theory. In this set-up, the vector multiplet $(\phi,\Lambda,D)$ appears very naturally. On the other hand, a purely field-theoretic derivation from the instantons of the $\nn=4$ theory would yield a different looking answer \cite{instrev} which does not refer to the moduli $\phi$, $\Lambda$ or $D$. This is possible because these variables play a superficially trivial and technical r\^ole in \eqref{Sp}. Since $D$ and $\Lambda$ appear only linearly, they are Lagrange multipliers and can be eliminated by imposing directly the associated ADHM constraints
%
%
via a $\delta$ function in the path integral measure.
Since $\phi$ appears only quadratically, it can be integrated out exactly, yielding an action $S^{(4)}$ for the usual reduced set of ADHM variables $(X,\bar\psi,q,\chi,\tilde q,\tilde\chi)$. The action $S^{(4)}$ is the standard quartic action for the ADHM moduli \cite{instrev}, corrected by terms coupling to the $\nn=4$ local fields coming from the diagrams of type d) in figure \ref{fig3}, as in \eqref{Sp31}. From this point of view, \eqref{pathint1} takes the field theoretic form
\be\label{pathintFT} \int\!\d\mu_{\text{b}}\d
\mu_{\text{ADHM}}\,
e^{-S_{\text{b}}-S^{(4)}}\, ,
\ee
where $\d\mu_{\text{ADHM}}$ is the well-known measure over the ADHM
moduli space \cite{instrev}.
Was then the multiplet $(\phi,\Lambda,D)$ a trivial artefact of the string approach? Or does it have a physical significance and, if yes, in what precise sense?

The planar diagrams associated with \eqref{pathintFT} can be divided
in two categories, of which typical representatives are depicted in figure \ref{fig4}. The diagram on the left, coming from the quartic couplings in $S^{(4)}$, looks like a tree of ``bubbles.'' Such diagrams are typical of the so-called vector models. The simplest vector model is a theory of $N$ scalar fields $\vec Q = (Q_{1},\cdots,Q_{N})$ with a quartic interaction 
\be\label{Lint} L_{\text{int}} = g\vec Q^{4}\, .\ee
The multi-loop Feynman diagrams are exactly as on the left of figure 
\ref{fig4} (with no dashed line). The technique to sum them up is well known
\cite{vectorrev}. One introduces an auxiliary field $\phi$ to rewrite the quartic interaction as
\be\label{Lint2} L_{\text{int}}' = -\phi^{2} + 2\sqrt{g}\phi\vec Q^{2}\, .\ee
Since $\phi$ appears only quadratically in \eqref{Lint2}, it can be integrated out exactly. This sets
\be\label{phiform} \phi = \sqrt{g}\vec Q^{2}\ee
and we find \eqref{Lint} back from \eqref{Lint2}. The trick to sum the bubble diagrams is instead to keep $\phi$ and to integrate out $\vec Q$, which also appears quadratically in \eqref{Lint2}.
The gaussian integration
yields an effective action for $\phi$ \emph{which is automatically proportional to $N$.} At large $N$, the composite variable \eqref{phiform} thus becomes classical and the physics is dominated by tree diagrams; these trees are precisely the tree-like bubbles of figure \ref{fig4}, drawn in a dual way in terms of $\phi$ \cite{vectorrev}. Of course there are also diagrams in the form of loops of bubbles, which are suppressed by powers of $1/N$. When the model is written in terms of $\phi$, these are associated with ordinary loop corrections.

This technique works for any large $N$ theory built from fields carrying a vector index. For example, both supersymmetric and non-supersymmetric models, which were a crucial source of inspiration for the present work, were studied in \cite{Neumann}. One always introduces auxiliary variables in such a way that the vector fields appear only quadratically in the action.
In our case, the ``vector'' index associated with the bubble diagrams in figure \ref{fig4} is the $\text{U}(N)$ index $f$ and the vector variables for this index in the path integral \eqref{pathintFT} are the moduli 
$(\tilde q^{f}, q_{f},\tilde\chi^{f},\chi_{f})$.
The required auxiliary variables are of course the moduli $(\phi,\Lambda,D)$. Indeed, when these variables are introduced, the action becomes quadratic in the vector moduli, see \eqref{Sp31}. Integrating out 
$\tilde q,q,\tilde\chi,\chi$ from \eqref{pathint1} then produces an effective action $\tilde S_{\text{eff}}$ which is automatically proportional to $N$, exactly as in the case of the simpler model \eqref{Lint}, \eqref{Lint2}. \emph{This implies that the fluctuations of the vector multiplet variables are suppressed in the large $N$ limit. In particular,}
\be\label{YAdef} Y_{A}= \ls^{2}\phi_{A}\ee
\emph{will be interpreted as the emergent space coordinates}. This is exactly the same mechanism which is responsible for the emergence of the bulk holographic coordinates in \cite{instbrit1}.

With the above interpretation in mind, let us thus perform the gaussian integration over $(q,\tilde q,\chi,\tilde\chi)$ in \eqref{pathint1}. We get
\be\label{pathint3} \int\!\d^{N^{2}}\!\!\mu_{\text{b}}
\,\d X\d\bar\psi\d\phi\d\Lambda\d D
\,\frac{\det ( \mathsf F + \mathsf V\mathsf B^{-1}
\mathsf W)}{\det\mathsf B}\, e^{-S_{\text b}- S_{\text{p},-1/-1}}
\, ,\ee
where the matrices $\mathsf B$, $\mathsf F$, $\mathsf V$ and $\mathsf W$ are of sizes $2NK\times 2NK$, $4NK\times 4NK$, $4NK\times 2NK$ and $2NK\times 4 NK$ respectively and are given explicitly by
\begin{align}\label{BFvwdef}\nonumber
&\mathsf B_{\alpha f i}^{\ \ \ \ \beta f' j} =
\frac{1}{2}\delta_{\alpha}^{\beta}
\bigl(\phi_{Ai}^{\ \ k}\delta_{f}^{f''} - \delta_{i}^{k}\hat\varphi_{Af}^{\ \ f''}\bigr)  \bigl(\phi_{Ak}^{\ \ j}\delta_{f''}^{f'} - \delta_{k}^{j}\hat\varphi_{Af''}^{\ \ f'}\bigr)+
\frac{i}{2}
\bigl(D_{\mu\nu i}^{\ \ \ j}\delta_{f}^{f'} - \delta_{i}^{j}\hat F_{\mu\nu f}^{+\ f'}\bigr)\sigma_{\mu\nu\alpha}^{\hphantom{\mu\nu\alpha}\beta}
\\\nonumber
&\mathsf F_{a fi\ b}^{\ \ \ \ \ f'j} = -\frac{1}{2}
\Sigma_{Aab}\bigl(\phi_{Ai}^{\ \ j}\delta_{f}^{f'}-\delta_{i}^{j}\hat\varphi_{Af}^{\ \ f'}\bigr)\\\nonumber
&\mathsf V_{afi}^{\ \ \ \alpha f' j}=\frac{1}{\sqrt{2}}\bigl(\Lambda^{\alpha\ j}_{\ ai}\delta_{f}^{f'} -
\hat\lambda^{\alpha\ f'}_{\ af}\delta_{i}^{j}\bigr)\\
&\mathsf W_{\alpha fi\ a}^{\ \ \ \ \ f'j}=\frac{1}{\sqrt{2}}
\bigl(\Lambda_{\alpha a i}^{\ \ \ j}\delta_{f}^{f'}-\delta_{i}^{j}\hat\lambda_{\alpha a f}^{\ \ \ f'}\bigr)\, .
\end{align}
The determinants in \eqref{pathint3} are $\text{U}(N)$ singlets. They depend on the fields of the $\nn=4$ theory living on the D3 branes only through gauge invariant local operators. If needed, the explicit form of these operators can be obtained straightforwardly from \eqref{BFvwdef} and \eqref{locexp}, for example by expanding the determinants in powers of the $\nn=4$ fields.

\subsection{\label{sub42} Matrix model diagrams}

The bubble diagrams of the previous section couple to the local $\nn=4$ fields and integrating over these fields amounts to summing up complicated planar diagrams as the one depicted on the right of figure \ref{fig4}, with the $\nn=4$ theory propagators and vertices decorating the loops in the bubbles. If we define
\be\label{tildeseffdef} e^{-\tilde S_{\text{eff}}(\phi,\Lambda,D)} =
\int\!\d^{N^{2}}\!\!\mu_{\text{b}}\frac{\det ( \mathsf F + \mathsf V\mathsf B^{-1}
\mathsf W)}{\det\mathsf B}\, e^{-S_{\text b}}
= \biggl\langle \frac{\det ( \mathsf F + \mathsf V\mathsf B^{-1}
\mathsf W)}{\det\mathsf B}\biggr\rangle_{\mathcal N=4}
\ee
then \eqref{pathint3} becomes
\be\label{pathint4}
\int\!\d X\d\bar\psi\d\phi\d\Lambda\d D\,
e^{-\tilde S_{\text{eff}}(\phi,\Lambda,D)-S_{\text p,-1/-1}(\phi,\Lambda,D,X,\bar\psi)}\, .\ee
We have succeeded in putting our partition function in a form that matches the right-hand side of \eqref{pathint2}! 

Anticipating slightly the next section, it is actually useful to improve the formula a bit further. The reason for this is that the light degrees of freedom living on a D-instanton correspond to the dimensional reduction of the ten dimensional $\nn=1$ vector multiplet down to zero dimension, which yields ten embedding matrix coordinates $Z$ and sixteen fermions
$\Psi$ in the adjoint of $\text{U}(K)$. In terms of the $\text{Spin}(4)\times\text{Spin(6)}\subset\text{Spin(10)}$ decomposition, this corresponds to 
\be\label{PsiZdef}
Z_{M}=(X_{\mu},Y_{A})\, ,\quad \Psi = (\psi_{\alpha a},\bar\psi^{\dot\alpha a})\, ,\ee
where $Y_{A}$ is defined by \eqref{YAdef} and 
\be\label{psidef}\psi_{\alpha a}=\ls^{2}\Lambda_{\alpha a}\, .\ee
There is no sign of the moduli $D$ and thus it is natural to integrate it out from \eqref{pathint4}. This is conceptually simple.  
As we have emphasized previously, the action $\tilde S_{\text{eff}}$ is always proportional to $N$; in our explicit formula \eqref{tildeseffdef}, this comes from the fact that the determinants involve products over the $\text{U}(N)$ indices. To leading order in $1/N$, integrating out $D$ thus simply amounts to solving the equation of motion for $D$ and plugging back the solution $\langle D_{\mu\nu}\rangle$ of this equation into the action.
Using \eqref{Sp1}, the equation of motion reads
\be\label{Deq} \frac{2i\pi}{\gs}\bigl[X_{\mu},X_{\nu}\bigr]_{\ i}^{+\ j}
+\frac{\partial \tilde S_{\text{eff}}}{\partial D_{\mu\nu i}^{\ \ \ j}}
= 0\, ,\ee
where the $+$ in the superscript of the commutator in \eqref{Deq} means that we consider the self-dual part of the commutator,
\be\label{dscommdef} \bigl[X_{\mu},X_{\nu}\bigr]^{+} = \frac{1}{2}\Bigl(
\bigl[X_{\mu},X_{\nu}\bigr] + \frac{1}{2}\epsilon_{\mu\nu\rho\sigma}
\bigl[X_{\rho},X_{\sigma}\bigr]\Bigr)\, .\ee
Overall, defining
\begin{multline}\label{Seffdef} S_{\text{eff}}=
\frac{\pi}{\gs\ls^{4}}\truK \Bigl\{
2i\ls^{4}\langle D_{\mu\nu}\rangle\bigl[X_{\mu},X_{\nu}\bigr] -
\bigl[X_{\mu},Y_{A}\bigr]\bigl[X_{\mu},Y_{A}\bigr]\\
-2\ls^{2}  \psi^{\alpha}_{\ a}\sigma_{\mu\alpha\dot\alpha}
\bigl[X_{\mu},\bar\psi^{\dot\alpha a} \bigr]
- \ls^{2}\bar\psi_{\dot\alpha}^{\ a}\Sigma_{Aab}
\bigl[Y_{A},\bar\psi^{\dot\alpha b}\bigr]\Bigr\} +
\tilde S_{\text{eff}}\bigl(Y,\psi,\langle D\rangle\bigr)\, ,
\end{multline}
we get the final holographic, or closed string, form for our path integral,
\be\label{pathint5}
\int\!\d X\d Y\d\psi\d\bar\psi\, e^{- S_{\text{eff}}}\, ,\ee
exactly as in the right-hand side of \eqref{pathint2}.

\subsection{\label{sub43} The state/geometry mapping}

The above construction can be generalized to associate an action $S_{\text{eff}}^{\rho}$ to any state $\rho$, pure or mixed, of the $\nn=4$ gauge theory. One simply has to replace the vacuum expectation value appearing in \eqref{tildeseffdef} by the expectation value in the state $\rho$. Since, as explained in the next section, a geometry can be associated 
to the action $S_{\text{eff}}^{\rho}$, this yields a mapping between the Hilbert space of the $\nn=4$ theory and type IIB supergravity backgrounds. 
For example, one can access in principle black hole geometries by computing \eqref{tildeseffdef} in the canonical ensemble at temperature $T$ and solving \eqref{Deq}. This construction can also be fruitfully generalized beyond the 
D$3$/D$(-1)$ system \cite{toappeara,toappearb,toappearc}.

If we focus on the conformal vacuum of the $\nn=4$ theory, in which $\langle\varphi\rangle = 0$, a remarkable simplification occurs and the expectation value in \eqref{tildeseffdef} can be evaluated effortlessly. This simplification comes from the fact that the local $\nn=4$ operators that appear in $\mathsf B$, $\mathsf F$, $\mathsf V$ and $\mathsf W$ are all evaluated at the same point, as was explained in section \ref{s3} after equation \eqref{locexp}. The expectation value we have to compute is thus
\emph{a one-point function of a gauge invariant operator} (the determinants in \eqref{tildeseffdef} are clearly $\text{U}(N)$ invariant). The result of such a computation is always trivial in a unitary conformal field theory and thus
\be\label{vevconformal}
\biggl\langle \frac{\det ( \mathsf F + \mathsf V\mathsf B^{-1}
\mathsf W)}{\det\mathsf B}\biggr\rangle_{\mathcal N=4} =
\frac{\det ( \tilde{\mathsf f} + \tilde{\mathsf v}\tilde{\mathsf b}^{-1}
\tilde{\mathsf w})}{\det\tilde{\mathsf b}}
\ee
where $\tilde{\mathsf b}$, $\tilde{\mathsf f}$, $\tilde{\mathsf v}$ and $\tilde{\mathsf w}$ and obtained from $\mathsf B$, $\mathsf F$, $\mathsf V$ and $\mathsf W$ in \eqref{BFvwdef} by setting the $\mathcal N=4$ local fields to zero. The $\text{U}(N)$ index structure then becomes trivial and the formula can be further simplified to
\be\label{vevconformal2}
\biggl\langle \frac{\det ( \mathsf F + \mathsf V\mathsf B^{-1}
\mathsf W)}{\det\mathsf B}\biggr\rangle_{\mathcal N=4} =
\biggl(
\frac{\det ( \mathsf f + \mathsf v\mathsf b^{-1}
\mathsf w)}{\det\mathsf b}\biggr)^{N}
\ee
where
\begin{align}\nonumber
&\mathsf b_{\alpha i}^{ \ \ \beta j} =\frac{1}{2}\delta_{\alpha}^{\beta}\bigl(
\phi_{A}\phi_{A}\bigr)_{i}^{\ j}
+\frac{i}{2}D_{\mu\nu i}^{\ \ \ j}\sigma_{\mu\nu\alpha}^{\hphantom{\mu\nu\alpha}\beta}\, ,\quad \mathsf f_{a i\ b}^{\ \ \ \ j} = -\frac{1}{2}
\Sigma_{Aab}\phi_{Ai}^{\ \ j}\ ,
\\\label{bfvwdef}
&\mathsf v_{ai}^{\ \ \alpha j}=
\frac{1}{\sqrt{2}}\Lambda^{\alpha\ j}_{\ ai}\ ,\quad
\mathsf w_{\alpha i\ a}^{\ \ \ \ j}=\frac{1}{\sqrt{2}}
\Lambda_{\alpha a i}^{\ \ \ j}\, .
\end{align}
In particular, \eqref{vevconformal2} makes manifest the fact that the effective action is proportional to $N$.

Of course, in general, the computation of the expectation value in \eqref{tildeseffdef}, or equivalently the sum over the planar diagrams of the type b) in figure \ref{fig4}, is very difficult. For instance, this will undoubtedly be the case if one wants to study black holes. However, let us emphasize that the emergence of space and the associated geometrical interpretation follow from \eqref{pathint5}, independently of our ability to compute \eqref{tildeseffdef} explicitly. On the other hand, the expectation value \eqref{tildeseffdef} encodes the details of the geometrical background (explicit metric and other fields), including the full set of $\alpha'$ corrections, as will become clear from the discussion in the next section.

Let us finally note that we have found strong evidence in \cite{toappeara,toappearb} for the non-renormalization of
the expectation value in \eqref{tildeseffdef} as well as its generalizations to other brane systems, if enough supercharges are preserved and independently of conformal invariance. 
This calls for further field theoretic study, but presently we shall stick to the simple conformal case for which the simple argument given above based on the triviality of one-point functions is sufficient.

\section{\label{s5} Reading off the geometry}

The right-hand side of the equality depicted in figure \ref{fig2} provides a simple interpretation of the effective action $S_{\text{eff}}$ appearing in \eqref{pathint2} or \eqref{pathint5}: it should correspond to the non-abelian action for $K$ D-instantons in an arbitrary type IIB background.
We are now going to explain how the full background can be read off unambiguously and quite straightforwardly from this action.

As explained above equation \eqref{PsiZdef}, the degrees of freedom in $S_{\text{eff}}$ correspond to matrix coordinates $Z_{M}$ for $M\leq 1\leq 10$ and associated fermions $\Psi$ in the adjoint representation of $\text{U}(K)$. For our purposes, it will be enough to consider the purely bosonic part of the action. We thus set 
\be\label{Psizero} \Psi = 0\ee
from now on. A convenient way to analyze the action is through its expansion around a diagonal configuration corresponding to a given space-time point $z$,
\be\label{Zexp} Z_{M} = z_{M} + \ls^{2}\epsilon_{M}\, .\ee
The expansion has the general form
\be\label{SDb}
S_{\text{eff}} = \sum_{n\geq 0} S_{\text{eff}}^{(n)}= \sum_{n\geq 0} \frac{1}{n!} \ls^{2n} c_{M_{1}\cdots M_{n}}(z)\tr\epsilon^{M_{1}}\cdots\epsilon^{M_{n}}\, .
\ee
The single trace structure is a consequence of the disk topology of the leading large $N\sim 1/\gs$ string diagrams on the right-hand side of figure \ref{fig2}. The coefficients $c_{M_{1}\cdots M_{n}}(z)$ can in principle be computed in terms of the non-trivial background by evaluating
diagrams with $n$ vertex operator insertions on the disk boundary, but this is notoriously difficult. Fortunately, using constraints coming from T-duality, Myers has been able to derive a form of the action which should be valid to leading order in $\alpha'$ for each coefficient and up to order five in the expansion \eqref{SDb} \cite{Myers}. This turns out to be precisely what we need to get the background.

Myers' action is the sum of Dirac-Born-Infeld and Chern-Simons terms,
\be\label{Mac1} S_{\text{eff}} = S_{\text{DBI}} + S_{\text{CS}}\, .\ee
For the case of D-instantons, the Dirac-Born-Infeld part is given by
\be\label{SBI} S_{\text{DBI}} = 2\pi\Str e^{-\Phi}\sqrt{\det 
\bigl(\delta_{MN} + i\ls^{2}[\epsilon_{M},\epsilon_{P}]
(G_{PN} + B_{PN})\bigr)}\, ,\ee
where $\Phi$, $G_{MN}$ and $B_{MN}$ are the usual dilaton, metric and Kalb-Ramond two-form of the Neveu-Schwarz sector. The fields are evaluated at the matrix point \eqref{Zexp} and should be expanded in powers of $\epsilon$. The determinant acts on the indices $M,N$ (not on the $\text{U}(K)$ indices of the matrices $\epsilon$). The Str is an appropriate symmetrized trace on the $\text{U}(K)$ indices whose precise definition is given in \cite{Myers} (and which should provide the correct ordering for the action up to order five in the expansion \eqref{SDb}, but not beyond). The Chern-Simons part of the action is given by
\be\label{SCS} S_{\text{CS}} = 2i\pi\Str e^{i\ls^{2}i_{\epsilon}i_{\epsilon}}\sum_{q\geq 0} C_{2q}\wedge e^{B}\vert_{0-\text{form}}\, ,\ee
where we keep only the 0-form part in the right-hand side, the $C_{2q}$ are the Ramond-Ramond forms and $i_{\epsilon}$ the inner product. It is tedious but completely elementary to use \eqref{SBI} and \eqref{SCS} to compute the $S_{\text{eff}}^{(n)}$ for $0\leq n\leq 5$. It is convenient to introduce the combination
\be\label{taudef}\tau = -C_{0} + ie^{-\Phi}\ee
to express the result, which reads
\begin{align}\nonumber
S_{\text{eff}}^{(0)} & = -2i\pi K\tau\\\nonumber
S_{\text{eff}}^{(1)} & = -2i\pi \ls^{2}\partial_{M}\tau\tr \epsilon_{M}\\
\nonumber
S_{\text{eff}}^{(2)} & = -i\pi\ls^{4}\partial_{M}\partial_{N}\tau\tr
\epsilon_{M}\epsilon_{N}\\\nonumber
S_{\text{eff}}^{(3)} & =\bigl(-\frac{i\pi}{3}\ls^{6}\partial_{M}\partial_{N}\partial_{P}\tau - 
2\pi\ls^{4}\partial_{[M}(\tau B - C_{2})_{NP]}\bigr)\tr\epsilon_{M}\epsilon_{N}\epsilon_{P} \\\nonumber
S_{\text{eff}}^{(4)} & =\bigl( -\frac{i\pi}{12}\ls^{8}
\partial_{M}\partial_{N}\partial_{P}\partial_{Q}\tau
- \frac{3\pi}{2}\ls^{6}\partial_{M}
\partial_{[N}(\tau B - C_{2})_{PQ]}\\\nonumber
& \hskip 4cm-\pi\ls^{4}e^{-\Phi}(G_{MP}G_{NQ} - G_{MQ}G_{NP})\bigr)
\tr\epsilon_{M}\epsilon_{N}\epsilon_{P}\epsilon_{Q}
\\\nonumber
S_{\text{eff}}^{(5)} & = \Bigl(-\frac{i\pi}{60}\ls^{10}\partial_{M}\partial_{N}\partial_{P}\partial_{Q}\partial_{R}\tau 
- \frac{\pi}{3}\ls^{8}\partial_{P}\partial_{Q}\partial_{R}(\tau B - C_{2})_{MN}\\\nonumber &\hskip 4cm
- \pi\ls^{6}\partial_{R}\bigl(e^{-\Phi}
(G_{MP}G_{NQ} - G_{MQ}G_{NP})\bigr)\\\label{Mexpand} &\hskip 2cm
-i\pi\ls^{6}\partial_{[M}(C_{4}+ C_{2}\wedge B - \frac{\tau}{2}
B\wedge B)_{NPQR]}\Bigr)
\tr\epsilon_{M}\epsilon_{N}\epsilon_{P}\epsilon_{Q}\epsilon_{R}\, .
\end{align}

The expressions for the coefficients $c_{M_{1}\cdots M_{n}}$, $0\leq n\leq 5$, that follow from \eqref{Mexpand} by taking the cyclic combinations of the factors in front of the traces, may seem very complicated. However, the structure of the expansion can be much clarified 
by organizing the calculation in terms of the symmetry properties of each term and studying the general consistency conditions that any action of the form \eqref{SDb} must satisfy. This analysis will be presented in \cite{toappear2}, but let us briefly comment on some of the simplest properties.

First of all, the coefficients can be decomposed into irreducible tensors with fixed symmetry properties, taking into account the cyclicity constraint 
\be\label{cyclic}c_{M_{1}\cdots M_{n}}=c_{M_{n}M_{1}\cdots M_{n-1}}\, .\ee
For example, the first non-trivial decomposition occurs at order three, where the cyclic tensor $c_{MNP}$ decomposes into totally symmetric and totally antisymmetric pieces,
\be\label{c3dec} c_{MNP} = c_{(MNP)} + c_{[MNP]}\, .\ee
At order four and five, one finds three and six irreducible pieces respectively. One then imposes the invariance of the action under the simultaneous shifts
\be\label{shiftsym} z \rightarrow z + \ls^{2}\delta\, ,\quad \epsilon\rightarrow\epsilon - \delta\, ,\ee
which is a direct consequence of the fact that 
$S_{\text{eff}}$ only depends on the combination \eqref{Zexp}. This invariance yields conditions of two types \cite{toappear2}. 

One type relates higher order
coefficients to derivatives of lower order coefficients. For example, one shows that the completely symmetrized coefficient is given by
\be\label{symrel} c_{(M_{1}\cdots M_{n})} = \partial_{M_{1}}\cdots
\partial_{M_{n}}c\, .\ee
This relation and its generalizations \cite{toappear2} explain many of the complicated derivative terms in \eqref{Mexpand} and reduce greatly the number of non-trivial independent coefficients to the following four, 

\begin{align}
\label{c0} &c = - 2i\pi\tau\\
\label{c3} &c_{[MNP]}  = -\frac{12\pi}{\ls^{2}}\partial_{[M}( \tau B - C_{2})_{NP]}\\
\label{c4} &c_{[MN][PQ]} = -\frac{18\pi}{\ls^{4}}e^{-\phi} 
\bigl(G_{MP}G_{NQ}-G_{MQ}G_{NP}\bigr)\\
\label{c5}
&c_{[MNPQR]}  = -\frac{120 i \pi}{\ls^{4}}
\partial_{[M}\bigl( C_{4} + C_{2}\wedge B - \frac{1}{2}\tau B\wedge B
\bigr)_{NPQR]}\, .
\end{align}

The second type of conditions are differential equations on the coefficients. Up to order five, this yields two new independent constraints \cite{toappear2},
\be\label{formcons} \partial_{[M}c_{NPQ]}=0\, ,\quad
\partial_{[M}c_{NPQRS]}=0\, .\ee
In other words, if one interprets $c_{[MNP]}$ and $c_{[MNPQR]}$ as the components of differential forms, then these forms must be closed.

Let us observe that even though the above constraints goes a long way in explaining the form of many terms in \eqref{Mexpand}, this is not the case for the fourth order coefficient \eqref{c4}. By construction, this coefficient must have precisely the same symmetries as the Riemann tensor, but this does not imply that it factorizes in terms of a second rank symmetric tensor $G$ as in \eqref{c4}. It is thus likely that $\alpha'$ corrections will violate this factorization property. This will be of no concern to us, because these $\alpha'$ corrections to Myers' formula will vanish when they are evaluated on the maximally supersymmetric $\AdSS$ background.\footnote{I would like to thank the referee for raising this point. Note that the vanishing of the $\alpha'$ corrections to Myers' action  for $\AdSS$ is similar to the vanishing of the $\alpha'$ corrections to the supergravity equations of motion in the same background. We have actually not been able to find a detailed derivation of this result in the literature. Our findings in the next Section can be interpreted as giving a strong hint that it is indeed valid.}

Up to the standard supergravity gauge transformations, there is a unique set of supergravity fields consistent with equations \eqref{c0}--\eqref{c5} for given coefficients $c$, $c_{[MNP]}$, $c_{[MN][PQ]}$ and $c_{[MNPQR]}$. All we have to do to reconstruct full the supergravity solution is thus to expand the action $S_{\text{eff}}$ at order five. This is actually quite fortunate, because the single trace prescription in Myers' action is known to be correct up to order five, but to fail at order six and higher. The aim of the next section will be to compute this expansion for the action \eqref{Seffdef}.

There is, however, an important subtlety associated with the use of the D-instanton action \cite{toappear2}. This subtlety is associated with the freedom to perform redefinitions of the matrix variables $Z_{M}$ in the action $S_{\text{eff}}$. Any redefinition $Z\mapsto Z'$ that preserves the single trace structure of $S_{\text{eff}}$ is a priori allowed and induces transformation laws $c_{M_{1}\cdots M_{n}}\mapsto c'_{M_{1}\cdots M_{n}}$ on the coefficients of the expansion \eqref{SDb}. Clearly, the physical information contained in $S_{\text{eff}}$ cannot depend on these transformations. As explained in details in \cite{toappear2}, for the case under study in the present paper, the transformations $c_{M_{1}\cdots M_{n}}\mapsto c'_{M_{1}\cdots M_{n}}$ are supergravity gauge transformations and thus the use of the D-instanton action to derive the background is perfectly consistent. In more general cases, as the one studied in \cite{toappeara}, a more detailed analysis is necessary to extract the physical information contained in $S_{\text{eff}}$.

\section{\label{s6} The solution of the model}

Taking into account \eqref{tildeseffdef}, \eqref{vevconformal2}, \eqref{bfvwdef} \eqref{YAdef} and \eqref{lambdadef}, the bosonic part of the action \eqref{Seffdef} reads, up to an irrelevant constant, 
\begin{multline}\label{Seffbose} S_{\text{eff}}=
\frac{4\pi^{2}N}{\ls^{4}\lambda}\truK \Bigl\{
2i\ls^{4}\langle D_{\mu\nu}\rangle\bigl[X_{\mu},X_{\nu}\bigr] -
\bigl[X_{\mu},Y_{A}\bigr]\bigl[X_{\mu},Y_{A}\bigr]\Bigr\}\\ +
N\ln\det\bigl(Y_{A}Y_{A}\otimes\mathbb I_{2\times 2} + i\ls^{4}\langle D_{\mu\nu}\rangle
\otimes\sigma_{\mu\nu}\bigr) - N \ln\det Y_{A}\otimes\Sigma_{A}\, .
\end{multline}
The expectation value $\langle D_{\mu\nu}\rangle$ is determined by 
\eqref{Deq} which, in our case, reads
\be\label{Deqexpl} \frac{8\pi^{2}}{\lambda}\bigl[X_{\mu},X_{\nu}\bigr]_{\ i}^{+\ j}
+\ls^{4}\bigl(Y_{A}Y_{A}\otimes\mathbb I_{2\times 2} + i\ls^{4}
\langle D_{\rho\sigma}\rangle
\otimes\sigma_{\rho\sigma}\bigr)^{-1\ \beta j}_{\ \alpha i}
\sigma_{\mu\nu\beta}^{\ \ \ \alpha} = 0\, .\ee
Following \eqref{Zexp}, with $Z_{M}=(X_{\mu},Y_{A})$, we set
\be\label{Zexp2} X_{\mu}= x_{\mu} + \ls^{2}\epsilon_{\mu}\, ,\quad
Y_{A} = y_{A} + \ls^{2}\epsilon_{A}\, .\ee
Our goal is to expand $S_{\text{eff}}$ in powers of $\epsilon$ up to the fifth order. 

Let us start by finding the expansion of $\langle D_{\mu\nu}\rangle$. We shall see that it is enough to compute this expansion at order three to get the expansion of $S_{\text{eff}}$ at order five. Using the identity
\be\label{Minv} \bigl( M+\delta M\bigr)^{-1} = M^{-1}-M^{-1}\delta M M^{-1} + \mathcal O\bigl((\delta M)^{2}\bigr)\, ,\ee
which is valid for any invertible matrix $M$ and small perturbation $\delta M$, together with the trace identity \eqref{tracesig4D}, we get
\be\label{Dexp1} \bigl\langle D_{\mu\nu}\bigr\rangle = \frac{4i\pi^{2}}{\lambda\ls^{4}} Y_{A}Y_{A}[\epsilon_{\mu},\epsilon_{\nu}]^{+}Y_{A}Y_{A} + 
\mathcal O\bigl(\langle D\rangle^{2}\bigr)\, .\ee
This shows that $\langle D\rangle$ is of order two in $\epsilon$ and thus that the terms $\mathcal O(\langle D\rangle^{2})$ are of order four. The order three formula we seek is thus immediately obtained from \eqref{Dexp1} by expanding $Y_{A}Y_{A}$ to linear order in $\epsilon$. Using a convenient six dimensional vector notation $(y_{A})=\vec y$, $(\epsilon_{A})=\vec\epsilon$, this yields
\be\label{Dexp2} \bigl\langle D_{\mu\nu}\bigr\rangle = \frac{4i\pi^{2}}{\lambda\ls^{4}}\Bigl( {\vec y\,}^{4}[\epsilon_{\mu},\epsilon_{\nu}]^{+} 
+ 2\ls^{2}{\vec y\,}^{2}\bigl( \vec y\cdot\vec\epsilon\, [\epsilon_{\mu},\epsilon_{\nu}]^{+} + [\epsilon_{\mu},\epsilon_{\nu}]^{+}\vec y\cdot\vec\epsilon\,\bigr)\Bigr) + \mathcal O\bigl(\epsilon^{4}\bigr)\, .\ee
Plugging this expression in \eqref{Seffbose} and expanding up to quadratic order in $\langle D\rangle$ by using the identity
\be\label{detexp} \ln\det (M+\delta M) = \ln\det M + \sum_{n\geq 1}
\frac{(-1)^{n+1}}{n}\tr (M^{-1}\delta M)^{n}\ee
we get
\begin{multline}\label{Seffa}S_{\text{eff}} = -\frac{N\ls^{8}}{2R^{4}}\truK\Bigl\{ 2[\epsilon_{\mu},\epsilon_{A}][\epsilon_{\mu},\epsilon_{A}] 
+
\frac{\vy^{4}}{R^{4}}[\epsilon_{\mu},\epsilon_{\nu}][\epsilon_{\mu},\epsilon_{\nu}]\Bigr\}\\
 -\frac{N\ls^{10}{\vec y\,}^{2}}{R^{8}}\truK\Bigl\{
2\vec y\cdot\vec\epsilon\, [\epsilon_{\mu},\epsilon_{\nu}][\epsilon_{\mu},\epsilon_{\nu}]+\epsilon_{\mu\nu\rho\sigma}\vec y\cdot\vec\epsilon\, [\epsilon_{\mu},\epsilon_{\nu}][\epsilon_{\rho},\epsilon_{\sigma}]\Bigr\}\\
+ 2N\ln\det \bigl(\vec y + \ls^{2}\vec\epsilon\,\bigr)^{2} - N\ln\det
\bigl(\vec y + \ls^{2}\vec\epsilon\,\bigr)\otimes\vec\Sigma
+\mathcal O\bigl(\epsilon^{6}\bigr)\, ,
\end{multline}
where we have defined
\be\label{Rdef} R^{4} = \frac{\ls^{4}\lambda}{4\pi^{2}}\,\cdotp\ee
We can finally expand the two remaining determinants using \eqref{detexp} and the trace identities \eqref{traceSigid},
\begin{multline}\label{bosdet}
2\ln\det \bigl(\vec y + \ls^{2}\vec\epsilon\,\bigr)^{2} = 2K\ln \vy^{2}
+ \frac{4\ls^{2}}{\vy^{2}}\truK \vec y\cdot\vec\epsilon
+ \frac{2\ls^{4}}{\vy^{4}}\truK\Bigl\{ \vy^{2}\ve^{2}- 2
\bigl(\vy\!\cdot\ve\bigr)^{2}\Bigr\}\\
+ \frac{16\ls^{6}}{\vy^{6}}\truK\Bigl\{-\frac{1}{4}\vy^{2}\ve^{2}\vy\!\cdot\ve + \frac{1}{3}\bigl(\vy\!\cdot\ve)^{3}\Bigr\}
+\frac{\ls^{8}}{\vy^{8}}\truK\Bigl\{ -\vy^{4}\ve^{4}+ 8\vy^{2}\ve^{2}
\bigl(\vy\!\cdot\ve\bigr)^{2}- 8 \bigl(\vy\!\cdot\ve\bigr)^{4}\Bigr\}\\
+ \frac{\ls^{10}}{\vy^{10}}\truK\Bigl\{ 4\vy^{4}\ve^{4}
\vy\!\cdot\ve-16 \vy^{2}\ve^{2}\bigl(\vy\!\cdot\ve\bigr)^{3}
+\frac{64}{5}\bigl(\vy\!\cdot\ve\bigr)^{5}\Bigr\}
+\mathcal O\bigl(\epsilon^{6}\bigr)
\end{multline}
and
\begin{multline}\label{fermdet}
\ln\det
\bigl(\vec y + \ls^{2}\vec\epsilon\,\bigr)\otimes\vec\Sigma = 
2K\ln \vy^{2}
+ \frac{4\ls^{2}}{\vec y^{2}}\truK \vec y\cdot\vec\epsilon
+ \frac{2\ls^{4}}{\vy^{4}}\truK\Bigl\{ \vy^{2}\ve^{2}- 2
\bigl(\vy\!\cdot\ve\bigr)^{2}\Bigr\}\\
+ \frac{16\ls^{6}}{\vy^{6}}\truK\Bigl\{-\frac{1}{4}\vy^{2}\ve^{2}\vy\!\cdot\ve + \frac{1}{3}\bigl(\vy\!\cdot\ve)^{3}\Bigr\}\\
+\frac{\ls^{8}}{\vy^{8}}\truK\Bigl\{ -2 \vy^{4}\ve^{4} +
8\vy^{2}\ve^{2}\bigl(\vy\!\cdot\ve\bigr)^{2}
- 8 \bigl(\vy\!\cdot\ve\bigr)^{4}+ \vy^{4}\epsilon_{A}\epsilon_{B}
\epsilon_{A}\epsilon_{B}\Bigr\}\\
+\frac{\ls^{10}}{\vy^{10}}\truK\Bigl\{ 4\vy^{4}\ve^{4}
\vy\!\cdot\ve-16 \vy^{2}\ve^{2}\bigl(\vy\!\cdot\ve\bigr)^{3}
+\frac{64}{5}\bigl(\vy\!\cdot\ve\bigr)^{5}\\-4\vy^{4}\vy\!\cdot\ve
\epsilon_{A}\epsilon_{B}\epsilon_{A}\epsilon_{B}+ 4\vy^{4}
\vy\!\cdot\ve\epsilon_{A}\ve^{2}\epsilon_{A} + \frac{4i}{5}
\vy^{4}\epsilon_{ABCDEF}\,y_{F}\,\epsilon_{A}\epsilon_{B}\epsilon_{C}\epsilon_{D}\epsilon_{E}\Bigr\}+\mathcal O\bigl(\epsilon^{6}\bigr)\, .
\end{multline}

Equations \eqref{bosdet} and \eqref{fermdet} imply that terms of order $\epsilon$, $\epsilon^{2}$ and $\epsilon^{3}$ exactly cancel in \eqref{Seffa},
\be\label{cancel} S_{\text{eff}}^{(1)}=S_{\text{eff}}^{(2)}=S_{\text{eff}}^{(3)}=0\, .\ee
From \eqref{taudef} and \eqref{Mexpand}, we see that
the vanishing $S_{\text{eff}}^{(1)}$ implies a constant dilaton and Ramond-Ramond zero form. Using \eqref{lambdadef} and introducing the usual gauge theory $\vartheta$ angle, we can write
\be\label{taures}\tau  = \frac{\vartheta}{2\pi} +
\frac{4i\pi N}{\lambda}\,\cdotp\ee
The vanishing of $S_{\text{eff}}^{(2)}$ does not yield any new information, consistently with the discussion in section \ref{s5}. On the other hand, \eqref{Mexpand} shows that the vanishing of $S_{\text{eff}}^{(3)}$ implies 
\be\label{S3van} \d (\tau B - C_{2}) = 0\ee
which, by taking the real and imaginary parts, yields the vanishing of
the Neveu-Schwarz and Ramond-Ramond three form field strengths
\be\label{BC2sol} \d B = H = 0\, ,\quad \d C_{2}=F_{3}=0\, .\ee

The first non-trivial terms in $S_{\text{eff}}$ are found at quartic and quintic orders. From \eqref{Seffa}, \eqref{bosdet} and \eqref{fermdet} we find
\begin{align}\label{Seff4} S_{\text{eff}}^{(4)} & =-\frac{N\ls^{8}}{2R^{4}}
\truK\Bigl\{ \frac{\vy^{4}}{R^{4}}[\epsilon_{\mu},\epsilon_{\nu}]
[\epsilon_{\mu},\epsilon_{\nu}] + \frac{R^{4}}{\vy^{4}}
[\epsilon_{A},\epsilon_{B}][\epsilon_{A},\epsilon_{B}] + 2
[\epsilon_{\mu},\epsilon_{A}][\epsilon_{\mu},\epsilon_{A}]\Bigr\}
\\\nonumber S_{\text{eff}}^{(5)} & = -\frac{2N\ls^{10}}{R^{8}}
\truK\Bigl\{\vy^{2}\vy\!\cdot\ve [\epsilon_{\mu},\epsilon_{\nu}]
[\epsilon_{\mu},\epsilon_{\nu}] - \frac{R^{8}}{\vy^{6}}
\vy\!\cdot\ve [\epsilon_{A},\epsilon_{B}][\epsilon_{A},\epsilon_{B}]
\Bigr\}\\
\label{Seff5}& \hskip .41cm
-\frac{4N\ls^{10}}{R^{8}}\truK\Bigl\{\epsilon_{\mu\nu\rho\sigma}
\vy^{2}\vy\!\cdot\ve \epsilon_{\mu}\epsilon_{\nu}
\epsilon_{\rho}\epsilon_{\sigma} +\frac{iR^{8}}{5\vy^{6}}
\epsilon_{ABCDEF}\,y_{F}\,\epsilon_{A}\epsilon_{B}\epsilon_{C}\epsilon_{D}\epsilon_{E}\Bigr\}\, .
\end{align}

Let us first analyze the quartic term. It should be compared with the corresponding term in \eqref{Mexpand} which, 
taking into account \eqref{taures} and \eqref{S3van}, reads
\be\label{Seff4M} S_{\text{eff}}^{(4)} = -\frac{N\ls^{8}}{2R^{4}}
G_{MP}G_{NQ}\truK [\epsilon_{M},\epsilon_{N}][\epsilon_{P},\epsilon_{Q}]\, .\ee
There is a perfect and unique match with \eqref{Seff4} (up to a trivial global sign of the metric which is fixed by positive-definiteness) corresponding to
\be\label{metsol} G_{\mu\nu} = \frac{\vy^{2}}{R^{2}}\delta_{\mu\nu}\, ,
\quad G_{AB} = \frac{R^{2}}{\vy^{2}}\delta_{AB}\, ,\quad
G_{\mu A} = 0\, .\ee
If we define the radial coordinate
\be\label{rdef} r^{2} = \vy^{2}\, ,\ee
the metric \eqref{metsol} reads
\be\label{AdSmet} \d s^{2} = \frac{r^{2}}{R^{2}}\d x_{\mu}\d x_{\mu}
+\frac{R^{2}}{r^{2}}\d r^{2} + R^{2}\d\Omega_{5}^{2}\, ,\ee
where $\d\Omega_{5}^{2}$ is the metric for the unit round five-sphere. We recognize the familiar $\AdSS$ metric, with the same correct radius $R$ defined in \eqref{Rdef} for the $\AdS$ and $\Sfive$ factors.

The quintic action in \eqref{Mexpand}, for constant $\tau$ and vanishing $H$ and $F_{3}$, contains two terms. Using \eqref{metsol}, it is straightforward to check that the term containing derivatives of
the metric precisely matches with the first line in \eqref{Seff5}. As explained in section \ref{s5}, this is nothing but a consistency check of our calculations. The non-trivial information comes from \eqref{c5}, which amounts to identifying the last line in \eqref{Seff5} with the remaining term in \eqref{Mexpand}. This term simplifies in our case to
\be\label{seff5msim} -\frac{i\pi\ls^{6}}{5} F_{5\, MNPQR}\truK
\epsilon_{M}\epsilon_{N}\epsilon_{P}\epsilon_{Q}\epsilon_{R}\, ,\ee
where $F_{5}=\d C_{4}$ is the Ramond-Ramond five-form field strength. This implies that
\begin{align}\nonumber (F_{5})_{\mu\nu\rho\sigma A} &= 
-\frac{4 i N\ls^{4}}{\pi R^{8}}\, \vy^{2}y_{A}\epsilon_{\mu\nu\rho\sigma}\, ,\\\label{F5sol}
(F_{5})_{ABCDE} &= \frac{4 N\ls^{4}}{\pi}\, \frac{y_{F}}{\vy^{6}}\epsilon_{ABCDEF}\, ,
\end{align}
with all the other components not related to \eqref{F5sol} by antisymmetry vanishing. Introducing the volume forms $\omega_{\AdS}$ and $\omega_{\Sfive}$ associated with the $\AdSS$ metric \eqref{AdSmet} and some choice of orientation,
\be\label{volumeforms} \omega_{\AdS} = \frac{\vy^{2}y_{A}}{R^{3}}\,
\d x_{1}\wedge\cdots\d x_{4}\wedge\d y_{A}\, ,\quad
\omega_{\Sfive} = \frac{1}{5!}\frac{R^{5}y_{F}}{\vy^{6}}\epsilon_{ABCDEF}\,\d y_{A}\wedge\cdots\wedge\d y_{E}\, ,\ee
we have
\be\label{F5sol2} F_{5} = \frac{4N\ls^{4}}{\pi R^{5}}\bigl(
\omega_{\Sfive} - i \omega_{\AdS}\bigr)\, . \ee
This five-form is precisely the correct solution of the type IIB supergravity field equations. In particular, it satisfies the euclidean version of the self-duality condition associated with \eqref{AdSmet},
\be\label{F5sd} \star F_{5} = -iF_{5}\, .\ee
Moreover, its flux
\be\label{Dirac} \int_{\Sfive}F_{5} = 4\pi^{2}N\ls^{4}\ee
is consistent with the string theoretic Dirac quantization condition for the $N$ D3-branes charge.
To find these detailed properties of type IIB supergravity and string theory emerging somewhat miraculously from our calculations is a rather non-trivial consistency check of the whole framework.

\section{\label{s7} Conclusion}

Let us summarize our main results:

--- We have described a framework from which the emergence of space can be explicitly derived. A basic idea is the separation of the set of planar diagrams in two categories as in figure \ref{fig4}. 
\emph{Space emerges from the tractable sum over the bubble diagrams.}

--- We have related the detailed geometrical properties of the emergent space to well-defined field theory correlators, as in \eqref{tildeseffdef}. This defines a precise mapping between states and geometries.

--- We have used our set-up to solve in full details the simplest example corresponding to the near-horizon geometry of a large number of D3 branes in type IIB string theory. We have been able to derive from first principles the full $\AdSS$ supergravity background, including the self-dual Ramond-Ramond five-form field strength. 

The point of view we have developed is a priori very general and can be applied to many different cases. For instance, models with no conformal invariance or no supersymmetry have also been studied successfully and will be published elsewhere \cite{toappeara,toappearb}. More generally, following the ideas presented in sections \ref{s2} and \ref{s4}, one can associate an emergent geometry to essentially any matrix field theory \cite{toappearc}.
It would also be interesting to revisit interesting previous works, like \cite{instbrit1,DPS}, in the light of our analysis.

An interesting aspect of our construction is that it yields a priori 
exact, finite $\alpha'$ non-abelian D-brane actions in non-trivial backgrounds. For example, the effective action defined by \eqref{Seffbose} and \eqref{Deqexpl} is supposed to be the exact non-abelian bosonic action for D-instantons in the $\AdSS$ background. The full supersymmetric version of the action can also be straightforwardly obtained from 
\eqref{Seffdef}, \eqref{vevconformal2}, \eqref{bfvwdef} and \eqref{Deq}. This opens the possibility to study various aspects of the non-abelian D-brane actions, like supersymmetry transformation laws and the action of diffeomorphisms, that are very difficult to discuss in general. 

Another nice feature is the possibility to study $1/N$ corrections, by summing the diagrams with loops of bubbles. These corrections are associated with the fluctuations of the (super) space coordinates $(X,\Psi)$ and correspond to loops in the theory defined by the effective action $S_{\text{eff}}\propto N$, see the right-hand side of \eqref{pathint2} or \eqref{pathint5}. Conceptually, having a fluctuating space and especially a fluctuating radial coordinate is satisfying. The radial coordinate is sometimes interpreted as a renormalization group scale, but such a scale does not have quantum fluctuations. This interpretation can thus be strictly valid only in the infinite $N$, classical gravity limit. In our framework, the radial coordinate is nothing but a field modulus, not surprisingly associated with the scale of the field configuration. Its quantum fluctuations follow directly from the quantum fluctuations of the field itself. 

As a final comment, we would like to emphasize a well-known, but possibly understated, fundamental consequence of the emergent space picture. The fluctuations of space and geometry are traditionally associated with the quantum corrections to a purely classical picture of gravity and thus, strictly speaking, to the genuine quantum gravity effects. This interpretation is misleading in models of emergent space. The microscopic, pre-geometric model we start with will always be treated quantum mechanically and the emergence of space and gravity are possible only as a consequence of strong quantum mechanical effects in this model, even when they look superficially classical. This could be the deepest lesson of this point of view: gravity and space itself are fundamentally quantum phenomena. This of course contradicts sharply the standard lore about the difficulties in quantum gravity, which is still advocated by a large fraction of the modern literature and which
presents gravity and quantum mechanics
as incompatible or at best hard to reconcile. If space emerges, there is really nothing to reconcile. Quite the contrary: we can find space and gravity only in a quantum mechanical framework. We believe that this tantalizing paradigm for quantum gravity could be universally accepted if only more effort would be devoted to the construction of tractable models.

\section{Acknowledgments}

I would like to thank Micha Moskovic, Antonin Rovai and particularly Jan Troost for useful discussions.

This work is supported in part by the belgian FRFC (grant 2.4655.07) and IISN (grants 4.4511.06 and 4.4514.08).

\begin{appendix}

\section{Conventions and identities}
\label{apA}

In this appendix, we provide definitions and identities used in the calculations presented in the main text.

\subsubsection*{Four dimensional matrices}

The Pauli matrices $\vec\sigma = (\sigma_{1},\sigma_{2},\sigma_{3})$ are defined as usual by
\be\label{Paulidef} \sigma_{1} = 
\begin{pmatrix} 0&1\\1&0\end{pmatrix}\, ,\quad \sigma_{2}=
\begin{pmatrix} 0&-i\\ i&0\end{pmatrix}\, ,\quad \sigma_{3}=\begin{pmatrix}
1&0\\0&-1\end{pmatrix}\, .\ee
We define $\sigma_{\mu\alpha\dot\alpha}$ and $\bar\sigma_{\mu}^{\dot\alpha\alpha}$, $1\leq\alpha,\dot\alpha\leq 2$, $1\leq\mu\leq 4$, by
\be\label{sigmudef}\sigma_{\mu}=(\vec\sigma,-i\mathbb I_{2\times 2})\, ,
\quad \bar\sigma_{\mu}=(-\vec\sigma,-i\mathbb I_{2\times 2})\, ,\ee
from which the euclidean four dimensional Dirac matrices, satisfying
\be\label{Cliff4D} \bigl\{\gamma_{\mu},\gamma_{\nu}\bigr\} = 2\delta_{\mu\nu}\, ,\ee
can be obtained,
\be\label{gamma4Ddef} \gamma_{\mu} = - i
\begin{pmatrix}
0 & \sigma_{\mu}\\ \bar\sigma_{\mu} & 0
\end{pmatrix}\, .
\ee
Weyl spinors $\lambda_{\alpha}$ and $\psi^{\dot\alpha}$ in the $(1/2,0)$ and $(0,1/2)$ representations of the
rotation group $\text{Spin}(4)=\text{SU}(2)_{+}\times
\text{SU}(2)_{-}$ transform under a four dimensional rotation parametrized by the antisymmetric matrix $\omega$, $\delta x_{\mu}= -\omega_{\mu\nu}x_{\nu}$, as
\be\label{spinorstl} \delta\lambda_{\alpha} = \frac{1}{2}\omega_{\mu\nu}
\sigma_{\mu\nu\alpha}^{\ \ \ \beta}\lambda_{\beta}\, ,\quad
\delta\psi^{\dot\alpha} = \frac{1}{2}\omega_{\mu\nu}\bar\sigma_{\mu\nu\ \dot\beta}^{\ \ \dot\alpha}\psi^{\dot\beta}\, ,\ee
where the generators of the rotation group are defined by
\be\label{gen4Ddef} \sigma_{\mu\nu}=\frac{1}{4}\bigl( \sigma_{\mu}\bar
\sigma_{\nu} - \sigma_{\nu}\bar\sigma_{\mu}\bigr)\, , \quad
\bar\sigma_{\mu\nu}=\frac{1}{4}\bigl( \bar\sigma_{\mu}
\sigma_{\nu} - \bar\sigma_{\nu}\sigma_{\mu}\bigr)\, .\ee
These symbols are self-dual and anti self-dual respectively,
\be\label{selfdualitysigma} \sigma_{\mu\nu}=\frac{1}{2}
\epsilon_{\mu\nu\rho\sigma}\sigma_{\mu\nu}\, ,\quad
\bar\sigma_{\mu\nu}=-\frac{1}{2}
\epsilon_{\mu\nu\rho\sigma}\bar\sigma_{\mu\nu}\, ,\ee
where $\epsilon_{\mu\nu\rho\sigma}$ is the completely antisymmetric tensor with $\epsilon_{1234}= + 1$. Useful trace identities are
\be\label{tracesig4D}\tr\sigma_{\mu\nu}\sigma_{\rho\sigma}
=-\frac{1}{2}\bigl( \delta_{\mu\rho}\delta_{\nu\sigma} - \delta_{\mu\sigma}
\delta_{\nu\rho} + \epsilon_{\mu\nu\rho\sigma}\bigr)\, ,\
\tr\bar\sigma_{\mu\nu}\bar\sigma_{\rho\sigma}
=-\frac{1}{2}\bigl( \delta_{\mu\rho}\delta_{\nu\sigma} - \delta_{\mu\sigma}
\delta_{\nu\rho} - \epsilon_{\mu\nu\rho\sigma}\bigr)\, .
\ee
Finally, indices on spinors $\lambda$ and $\bar\lambda$ can be raised or lowered as usual,
\be\label{raiselower}\lambda^{\alpha}=\epsilon^{\alpha\beta}\lambda_{\beta}\, ,\ \lambda_{\alpha} = \epsilon_{\alpha\beta}\lambda^{\beta}\, ,\
\psi^{\dot\alpha}=\epsilon^{\dot\alpha\dot\beta}\psi_{\dot\beta}\, ,\ \psi_{\dot\alpha} = \epsilon_{\dot\alpha\dot\beta}\psi^{\dot\beta}\, ,\ee
for completely antisymmetric two-dimensional symbols $\epsilon$ defined by $\epsilon^{12}=-\epsilon_{12}=+1$.

\subsubsection*{Six dimensional matrices}

We define
\begin{align}\nonumber
&\Sigma_{1}=
\begin{pmatrix} 0&-1&0&0\\1&0&0&0\\0&0&0&1\\0&0&-1&0\end{pmatrix}\, ,\
\Sigma_{2}=
\begin{pmatrix} 0&-i&0&0\\i&0&0&0\\0&0&0&-i\\0&0&i&0\end{pmatrix}\, ,\
\Sigma_{3}=
\begin{pmatrix} 0&0&-1&0\\0&0&0&-1\\1&0&0&0\\0&1&0&0\end{pmatrix}\, ,\\
\label{Sigmadef6D}
&\Sigma_{4}=
\begin{pmatrix} 0&0&-i&0\\0&0&0&i\\i&0&0&0\\0&-i&0&0\end{pmatrix}\, ,\
\Sigma_{5}=
\begin{pmatrix} 0&0&0&-1\\0&0&1&0\\0&-1&0&0\\1&0&0&0\end{pmatrix}\, ,\
\Sigma_{6}=
\begin{pmatrix} 0&0&0&-i\\0&0&-i&0\\0&i&0&0\\i&0&0&0\end{pmatrix}\, ,
\end{align}
and
\be\label{Sigmabardef6D} \bar\Sigma_{A}= \Sigma_{A}^{\dagger}\, .\ee
In particular we have
\be\label{relSigma} \bar\Sigma_{A}^{ab}= \frac{1}{2}\epsilon^{abcd}
\Sigma_{Acd}\, ,\quad 
\Sigma_{Aab}= \frac{1}{2}\epsilon_{abcd}\bar\Sigma_{A}^{cd}\ee
where the $\epsilon$s are completely antisymmetric symbols with $\epsilon_{1234}=\epsilon^{1234}=+1$.
Euclidean six dimensional Dirac matrices, satisfying
\be\label{Cliff6D} \bigl\{\Gamma_{A},\Gamma_{B}\bigr\} = 2\delta_{AB}\, ,\ee
can then be defined by
\be\label{gamma6Ddef} \Gamma_{A} = 
\begin{pmatrix}
0 & \Sigma_{A}\\ \bar\Sigma_{A} & 0
\end{pmatrix}\, .
\ee
Weyl spinors $\lambda_{a}$ and $\psi^{a}$ in the $\mathbf 4$ and $\mathbf{\bar 4}$ representations of the
rotation group $\text{Spin}(6)=\text{SU}(4)$ transform under a six dimensional rotation parametrized by the antisymmetric matrix $\Omega$, $\delta x_{A}= -\Omega_{AB}x_{B}$, as
\be\label{spinors6Dtl} \delta\lambda_{a} = -\frac{1}{2}\Omega_{AB}
\Sigma_{ABa}^{\ \ \ \ b}\lambda_{b}\, ,\quad
\delta\psi^{a} = -\frac{1}{2}\Omega_{AB}\bar\Sigma_{AB\ b}^{\ \ \ a}
\psi^{b}\, ,\ee
where the generators of the rotation group are defined by
\be\label{gen6Ddef} \Sigma_{AB}=\frac{1}{4}\bigl( \Sigma_{A}\bar
\Sigma_{B} - \Sigma_{B}\bar\Sigma_{A}\bigr)\, , \quad
\bar\Sigma_{AB}=\frac{1}{4}\bigl( \bar\Sigma_{A}
\Sigma_{B} - \bar\Sigma_{B}\Sigma_{A}\bigr)\, .\ee
If $\vec v = (v_{A})_{1\leq A\leq 6}$ is a six-dimensional vector, it is not difficult to check that
\begin{align}\label{detid}\det\bigl( v_{A}\Sigma_{A}\bigr) = (v_{A}v_{A})^{2}=\vec v^{4}\, ,\\ \label{inverseid}
\bigl(v_{A}\Sigma_{A}\bigr)^{-1}= \frac{v_{A}\bar\Sigma_{A}}{\vec v^{2}}\,\cdotp\end{align}
Trace identities that we use in section \ref{s5} of the main text can then be straightforwardly derived,
\begin{align}\nonumber
& v_{A'}\tr\bar\Sigma_{A'}\Sigma_{A} = 4 v_{A}\\\nonumber
& v_{A'}v_{B'}\tr\bar\Sigma_{A'}\Sigma_{A}\bar\Sigma_{B'}\Sigma_{B} =
8\, v_{A}v_{B} - 4\,\vec v^{2}\delta_{AB}\\\nonumber
& v_{A'}v_{B'}v_{C'}\tr\bar\Sigma_{A'}\Sigma_{A}\bar\Sigma_{B'}\Sigma_{B}
\bar\Sigma_{C'}\Sigma_{C}= 16\, v_{A}v_{B}v_{C}- 4\,\vec v^{2}\bigl(
v_{A}\delta_{BC}+v_{B}\delta_{AC}+v_{C}\delta_{AB}\bigr)\\\nonumber
&v_{A'}v_{B'}v_{C'}v_{D'}\tr\bar\Sigma_{A'}\Sigma_{A}\bar\Sigma_{B'}\Sigma_{B}\bar\Sigma_{C'}\Sigma_{C}\bar\Sigma_{D'}\Sigma_{D}=
32\, v_{A}v_{B}v_{C}v_{D} \\\nonumber & 
\hskip .6cm+ 4\, \vec v^{4}\bigl(\delta_{AB}\delta_{CD}
- \delta_{AC}\delta_{BD}+\delta_{AD}\delta_{BC}\bigr) - 8\,\vec v^{2}
\bigl(v_{A}v_{B}\delta_{BC}+\text{circular permutations}\bigr)\\
\nonumber &
v_{A'}v_{B'}v_{C'}v_{D'}v_{E'}\tr\bar\Sigma_{A'}\Sigma_{A}\bar\Sigma_{B'}\Sigma_{B}\bar\Sigma_{C'}\Sigma_{C}\bar\Sigma_{D'}\Sigma_{D}
\bar\Sigma_{E'}\Sigma_{E} =\\\nonumber&
\hskip 3cm \hphantom{+}64\, v_{A}v_{B}v_{C}v_{D}v_{E} - 16\, \vec v^{2}
\bigl( v_{A}v_{B}v_{C}\delta_{DE}+ \text{circular\ permutations}\bigr) \\\nonumber& \hskip 3cm
+ 4\,\vec a^{4}\bigl[ v_{A}(\delta_{BC}\delta_{DE}
-\delta_{BD}\delta_{CE}+\delta_{BE}\delta_{CD})
+ \text{circular permutations}\bigr]\\\label{traceSigid} &\hskip 3cm
+4 i\, \vec a^{4}a_{F}\epsilon_{ABCDEF}\, ,
\end{align}
where $\epsilon_{ABCDEF}$ is the six-dimensional completely antisymmetric tensor defined by $\epsilon_{123456}=+1$.

\subsubsection*{Indices and transformation laws}

See table 1.

\begin{table}
\be\nonumber
\begin{matrix}
& \text{Spin}(4)  & 
 \text{SU}(4) & \text{U}(N) & \text{U}(K)\\
\hline 
\alpha, \beta, ... \ \text{(upper or lower)}& (1/2,0) & \mathbf 1 & \mathbf 1&\mathbf 1 \\
\dot\alpha, \dot\beta, ... \ \text{(upper or lower)}& (0,1/2) & \mathbf 1 & \mathbf 1&\mathbf 1 \\
\mu, \nu, ... & (1/2,1/2) & \mathbf 1 & \mathbf 1&\mathbf 1 \\
a, b, ... \ \text{(lower)}& (0,0) & \mathbf 4 & \mathbf 1&\mathbf 1 \\
a, b, ... \ \text{(upper)}& (0,0) & \mathbf{\bar 4} & \mathbf 1&\mathbf 1 \\
A, B, ... & (0,0) & \mathbf{6} & \mathbf 1&\mathbf 1 \\
f, f', ... \ \text{(lower)}& (0,0) & \mathbf 1 & \mathbf N&\mathbf 1 \\
f, f', ... \ \text{(upper)}& (0,0) & \mathbf 1 & \mathbf{\bar N}&\mathbf 1 \\
i, j, ... \ \text{(lower)}& (0,0) & \mathbf 1 & \mathbf 1&\mathbf K \\
i, j, ... \ \text{(upper)}& (0,0) & \mathbf 1 & \mathbf 1&\mathbf{\bar K}\\
a_{\mu f}^{\ \ f'} & (1/2,1/2) & \mathbf 1 & \textbf{Adj} & \mathbf 1\\
\varphi_{A f}^{\ \ f'} & (0,0) & \mathbf 6 & \textbf{Adj} & \mathbf 1\\
\lambda_{\alpha a f}^{\ \ \ f'} & (1/2,0) & \mathbf 4 & \textbf{Adj} & \mathbf 1\\
\bar\lambda^{\dot\alpha a f'}_{\ \ f} & (0,1/2) & \mathbf{\bar 4} & \textbf{Adj} & \mathbf 1\\
X_{\mu i}^{\ \ j}=\ls^{2}A_{\mu i}^{\ \ j} & (1/2,1/2) & \mathbf 1 & \mathbf 1 & \textbf{Adj}\\
Y_{A i}^{\ \ j}=\ls^{2}\phi_{A i}^{\ \ j} & (0,0) & \mathbf 6 & \mathbf 1& \textbf{Adj} \\
\psi_{\alpha a i}^{\ \ \ j} =\ls^{2}\Lambda_{\alpha a i}^{\ \ \ j} & (1/2,0) & \mathbf 4 & \mathbf 1  & \textbf{Adj}\\
\bar\psi^{\dot\alpha a j}_{\ \ i} =\ls^{2}\bar\Lambda^{\dot\alpha a j}_{\ \ i} & (0,1/2) & \mathbf{\bar 4} & \mathbf 1
& \textbf{Adj} \\
D_{\mu\nu i}^{\ \ \ j} & (1,0) & \mathbf 1 & \mathbf 1 & \textbf{Adj}\\
q_{\alpha f i} & (1/2,0) & \mathbf 1 & \mathbf N  & \textbf{K}\\
\tilde q^{\alpha f i} & (1/2,0) & \mathbf 1 & \mathbf{\bar N}  & 
\mathbf{\bar K}\\
\chi^{a}_{\ f i} & (0,0) & \mathbf{\bar 4} & \mathbf{N}  & 
\mathbf{K}\\
\tilde\chi^{afi} & (0,0) & \mathbf{\bar 4} & \mathbf{\bar N}  & 
\mathbf{\bar K}
\end{matrix}
\ee
\caption{\label{indices}Conventions for the transformation laws of indices, fields and moduli. For maximum clarity, we have indicated all the indices associated to each field or modulus, whereas in the main text the gauge $\text{U}(N)$ and $\text{U}(K)$ indices are usually suppressed. The representations of $\text{Spin}(4)=\text{SU}(2)_{+}\times
\text{SU}(2)_{-}$ are indicated according to the spin in each $\text{SU}(2)$ factor. The $(1/2,1/2)$ of $\text{SU}(2)_{+}\times
\text{SU}(2)_{-}$ and the $\mathbf 6$ of $\text{SU}(4)=\text{Spin}(6)$ correspond to the fundamental representations of $\text{SO}(4)$ and $\text{SO}(6)$ respectively.}   
\end{table}

\section{Some supersymmetry transformation laws}
\label{apB}

We indicate for completeness in this appendix the supersymmetry transformation laws associated with the eight supercharges $Q^{\alpha a}$ 
that leave invariant the D$3$/D$(-1)$ system considered in the main text. We use the traditional notation $\delta\Phi = \xi_{\alpha a}Q^{\alpha a}\cdot
\Phi$ for the supersymmetry variation of an arbitrary field $\Phi$ with supersymmetry parameter $\xi$.

The action \eqref{Sb} is invariant under
\begin{align} \delta a_{\mu} & = -i\xi^{\alpha}_{\ a}\sigma_{\mu\alpha\dot\alpha}\bar\lambda^{\dot\alpha a}\\
\delta\varphi_{A}&=-i\xi^{\alpha}_{\ a}\bar\Sigma_{A}^{ab}\lambda_{\alpha b}
\\
\delta\lambda_{\alpha a}&=-\xi_{\beta a}\sigma_{\mu\nu\alpha}^{\ \ \ \beta}
F_{\mu\nu} - i \xi_{\alpha b}\bar\Sigma_{AB\ a}^{\ \ \ b}[\varphi_{A},\varphi_{B}]\\
\delta\bar\lambda^{\dot\alpha a} &= \xi_{\beta b}\bar\sigma_{\mu}^{\dot\alpha\beta}\bar\Sigma_{A}^{ba}\nabla_{\mu}\varphi_{A}\, .
\end{align}
This action is of course also invariant under similar transformations generated by the supercharges $\bar Q_{\dot\alpha a}$ which are broken in the presence of the D-instantons. By dimensional reduction, we obtain the transformation laws that leave \eqref{Sp1bis} invariant,
\begin{align} \delta A_{\mu} & = -i\xi^{\alpha}_{\ a}\sigma_{\mu\alpha\dot\alpha}\bar\Lambda^{\dot\alpha a}\\
\delta\phi_{A}&=-i\xi^{\alpha}_{\ a}\bar\Sigma_{A}^{ab}\Lambda_{\alpha b}
\\
\delta\Lambda_{\alpha a}&=-i\xi_{\beta a}\sigma_{\mu\nu\alpha}^{\ \ \ \beta}
[A_{\mu},A_{\nu}] - i \xi_{\alpha b}\bar\Sigma_{AB\ a}^{\ \ \ b}[\phi_{A},\phi_{B}]\\
\delta\bar\Lambda^{\dot\alpha a} &= i\xi_{\beta b}\bar\sigma_{\mu}^{\dot\alpha\beta}\bar\Sigma_{A}^{ba}[A_{\mu},\phi_{A}]\, .
\end{align}
Taking the $\ls\rightarrow 0$ limit with $X_{\mu}$ and $\bar\psi^{\dot\alpha a}$ defined in \eqref{Xmudef} and \eqref{barpsidef} fixed, and introducing the self-dual auxiliary field $D_{\mu\nu}$, we get the transformation laws that leave \eqref{Sp1} invariant,
\begin{align} \delta X_{\mu} & = -i\xi^{\alpha}_{\ a}\sigma_{\mu\alpha\dot\alpha}\bar\psi^{\dot\alpha a}\\
\delta\bar\psi^{\dot\alpha a} &= i\xi_{\beta b}\bar\sigma_{\mu}^{\dot\alpha\beta}\bar\Sigma_{A}^{ba}[X_{\mu},\phi_{A}]\\
\delta\phi_{A}&=-i\xi^{\alpha}_{\ a}\bar\Sigma_{A}^{ab}\Lambda_{\alpha b}
\\
\delta\Lambda_{\alpha a}&=-\xi_{\beta a}\sigma_{\mu\nu\alpha}^{\ \ \ \beta}
D_{\mu\nu} - i \xi_{\alpha b}\bar\Sigma_{AB\ a}^{\ \ \ b}[\phi_{A},\phi_{B}]\\
\delta D_{\mu\nu}&= -\xi^{\alpha}_{\ a}\sigma_{\mu\nu\alpha}^{\ \ \ \beta}
\bar\Sigma_{A}^{ab}[\phi_{A},\Lambda_{\beta b}]\, .
\end{align}
The action \eqref{Sp31t} is invariant under
\begin{align}
\delta q_{\alpha} & = -i\sqrt{2}\,\xi_{\alpha a}\chi^{a}\\
\delta\chi^{a} & = i\sqrt{2}\,\xi^{\alpha}_{\ b}\bar\Sigma_{A}^{ab}\phi_{A}
q_{\alpha}\\
\delta \tilde q^{\alpha} & = -i\sqrt{2}\,\xi^{\alpha}_{\ a}\tilde\chi^{a}\\
\delta\tilde\chi^{a} & = -i\sqrt{2}\,\xi_{\alpha b}\bar\Sigma_{A}^{ab}\phi_{A}\tilde q^{\alpha}\, .
\end{align}
Let us note that the triplet $(\phi,\Lambda,D)$ transforms as an off-shell vector multiplet of six dimensional $\nn=1$ supersymmetry (or four dimensional $\nn=2$ supersymmetry), whereas $(X,\bar\psi)$ and $(q,\tilde q,\chi,\tilde\chi)$ transform as a $\text{U}(K)$ adjoint and $N$ flavors of $\text{U}(K)$ fundamentals hypermultiplets respectively.

\end{appendix}
\end{document}